\documentclass{article}
\usepackage{graphicx}
\usepackage{biblatex}
\usepackage{subcaption}
\usepackage{svg}
\usepackage{amsmath}
\usepackage{booktabs}
\usepackage{hyperref}
\usepackage[toc,page]{appendix}
\usepackage{multirow}
\usepackage{float}
\usepackage{array}
\usepackage{siunitx}
\usepackage{longtable}
\usepackage[table]{xcolor}
\definecolor{blue}{HTML}{6087C6}
\definecolor{green}{HTML}{8BBD87}
\definecolor{purple}{HTML}{976ADF}
\definecolor{red}{HTML}{E77A7A}
\usepackage{authblk}
\usepackage{tabularx}
\usepackage[a4paper, margin=3cm]{geometry}
\usepackage{enumitem}
\newcolumntype{C}[1]{>{\centering\arraybackslash}p{#1}}

\title{Representation learning of human cortical folding to reveal long lasting neurodevelopmental signatures}
\author[1]{Julien Laval$^{+}$}
\author[2]{Robin Guiavarch$^{*}$}
\author[1]{Antoine Dufournet$^{*}$}
\author[1]{Racim Menasria}
\author[1]{Barthélémy Drabczuk}
\author[1]{Cristobal Mendoza}
\author[1]{Saeb Tounsi}
\author[1]{Chikh Abdelghani Baroud}
\author[3]{Merieme Bourenane}
\author[4]{Vanessa Troiani}
\author[5,6]{William Snyder}
\author[7]{Marisa A Patti}
\author[8]{Mylène Moyal}
\author[8]{Marion Plaze}
\author[9,10]{Arnaud Cachia}
\author[11]{Federica Santacroce}
\author[11]{Giorgia Committeri}
\author[12]{Claire Cury}
\author[13]{Kevin De Matos}
\author[13]{Olivier Colliot}
\author[13]{Zhong Yi Sun}
\author[1]{Clara Fischer}
\author[1]{Vincent Frouin}
\author[2]{Pietro Gori}
\author[1]{Denis Rivière}
\author[1]{Joël Chavas$^{\dag}$}
\author[1]{Jean-François Mangin$^{\dag}$}

\affil[1]{Université Paris-Saclay, CEA, CNRS, Neurospin, UMR 9027 BAOBAB, GAIA Laboratory, Gif-sur-Yvette, France}
\affil[2]{LTCI, Institut Polytechnique de Paris, Télécom Paris, Palaiseau, France}
\affil[3]{Maison de la simulation, Université Paris-Saclay, CEA, Gif-sur-Yvette,
France}
\affil[4]{Department of Developmental Medicine, Geisinger, Lewisburg, USA}
\affil[5]{National Institute of Mental Health Intramural Research Program, Bethesda, USA}
\affil[6]{Department of Psychiatry, University of Cambridge, Cambridge, UK}
\affil[7]{A.J. Drexel Autism Institute, Drexel University, USA}
\affil[8]{GHU Paris, IPNP, INSERM U1266, Paris, France}
\affil[9]{Université Paris Cité, LaPsyDÉ, CNRS, Paris F-75005, France}
\affil[10]{Institut Universitaire de France}
\affil[11]{Department of Psychology, ITAB – Institute for Advanced Biomedical Technologies, CARES – Centre for Disability, Rehabilitation and Sports Medicine, University G. d’Annunzio of Chieti-Pescara, Chieti, Italy}
\affil[12]{Univ Rennes, Inria, CNRS, Inserm - IRISA UMR 6074, Empenn ERL U 1228, Rennes, France}
\affil[13]{Sorbonne Université, Institut du Cerveau – Paris Brain Institute - ICM, CNRS, Inria, Inserm, AP-HP, Hôpital de la Pitié-Salpêtrière, F-75013, Paris, France}
\date{}

\begin{document}

\maketitle
\begingroup
\renewcommand\thefootnote{*}
\footnotetext{These authors contributed equally to this work.}
\renewcommand\thefootnote{\dag}
\footnotetext{These authors equally supervised this work.}
\renewcommand\thefootnote{+}
\footnotetext{E-mail: \texttt{julien.laval@cea.fr}}
\endgroup
\section{Abstract}

The human brain folds in utero, primarily during late gestation. Shortly after birth, cortical folding patterns are established and remain stable thereafter, making them promising early neurodevelopmental markers. Yet it is unclear whether the representations given by current neuroimaging foundation models capture cortical folding variability. Here, we introduce Champollion, a self-supervised learning framework that learns interpretable local representations of cortical folding from structural MRI. Optimized on representative folding-related tasks, Champollion accurately captures known folding patterns across cortical regions and external datasets. In a comprehensive benchmark, it consistently outperforms neuroimaging and general-purpose foundation models. Furthermore, Champollion reveals richer genetic associations than conventional morphometric descriptors and identifies localized folding signatures associated with incomplete hippocampal inversion, prematurity, and maternal smoking. These results establish cortical folding as a rich and largely untapped source of neurodevelopmental information, and Champollion provides a unified framework for discovering, localizing and interpreting long lasting cortical folding signatures.

\section{Introduction}

The human cerebral cortex is highly folded, consisting of ridges (gyri) separated by grooves (sulci). Cortical folding emerges primarily during the second half of gestation and reaches near-adult levels shortly after birth~\cite{De_Vareilles2023-ri}. Although sulci exhibit substantial inter-individual shape variability~\cite{Mangin2016-je}, this variability is not random. Prior to folding, spatial gradients of gene expression in the developing cortex closely match the future layout of sulci and gyri, indicating a genetic contribution to folding~\cite{De_Juan_Romero2017-lr}. Folding patterns arise from a complex interplay between gene expression and mechanical constraints that regulate neuronal proliferation, migration, and differential cortical expansion~\cite{Llinares-Benadero2019-jf}. These patterns also partially predict the underlying cytoarchitecture of the cortex~\cite{Fischl2008-nj}.

Unlike scalar descriptors of cortical morphology, such as sulcal depth or gyrification index (the magnitude of the convolutions), which evolve throughout development and aging~\cite{gyrification_1,gyrification_2}, the topology and shape of cortical folds remain remarkably stable once established~\cite{cachia_longitudinal_2016,tissier_sulcal_2018}. The early emergence and lifelong stability of these folding patterns make them particularly attractive for studying early neurodevelopmental mechanisms underlying brain function and pathology.

Consistent with this view, folding patterns have been linked to a range of behavioral and clinical phenotypes~\cite{voorhies_cognitive_2021,liang_cortical_2025}. For example, hemispheric asymmetries of topological features in the anterior cingulate cortex and inferior frontal cortex have been associated with inhibitory control~\cite{tissier_sulcal_2018}. Alterations in folding patterns have also been reported in psychiatric disorders, including schizophrenia~\cite{Yucel2002-kt,Provost2003-ux,Fujiwara2007-bo,Nakamura2007-pe,Penttila2008-pm,Cachia2008-go,Plaze2011-mk,garrison_paracingulate_2015,rollins_evidence_2020,Wu2025-iz}. Among these, different orbitofrontal sulcal pattern distributions have been observed in patients and controls~\cite{Nakamura2007-pe}, and the perceived localization of auditory hallucinations (internal versus external) has been linked to the shape of the right superior temporal sulcus~\cite{Plaze2011-mk}. However, most studies rely on manual annotations or a small number of handcrafted descriptors, limiting the systematic exploration of folding variability at scale.

Recent advances in large-scale brain magnetic resonance imaging (MRI) datasets and self-supervised learning (SSL) have transformed feature extraction in neuroimaging. Rather than relying on predefined anatomical measurements, SSL learns representations directly from imaging data, producing features that transfer across downstream tasks. This paradigm has fueled the emergence of increasingly general neuroimaging foundation models~\cite{dufumier_exploring_2024,barbano2025anatomical,cox2024brainsegfounder,xu2025threedino}. However, whether such generic representations adequately capture cortical folding, a highly variable signal whose developmental relevance is not explicit in the imaging data alone, remains largely unexplored.

Here, we introduce Champollion, an SSL framework to characterize cortical folding variability. For both consistency with previous studies on folding patterns and better interpretability, Champollion learns at a local scale, from regional crops of brain MRIs preprocessed for cortical folding extraction. For a given region of interest (ROI), typically centered on one sulcus or a small set of neighboring sulci, Champollion learns a low-dimensional representation that captures the local folding shape based on $\sim42{,}000$ samples from the UK Biobank~\cite{bycroft_uk_2018}.

We first show that Champollion learns spatially grounded representations that accurately capture diverse cortical folding patterns. We then show that these representations consistently outperform existing alternatives, including handcrafted morphometric descriptors, previous self-supervised approaches, and a broad range of recent vision foundation models trained on either natural images or brain images, demonstrating that current foundation models largely overlook cortical folding variability. Finally, we show that Champollion's representations capture biologically meaningful information beyond classical morphometry, enabling richer genetic associations and the discovery of subtle yet clinically relevant folding patterns, exploring increasingly challenging tasks: incomplete hippocampal inversion, great prematurity, and maternal smoking exposure during pregnancy.

\section{Results}

\subsection{Champollion learns fine-grained representations of cortical folding patterns}

\begin{figure}[H]
\centering
\includegraphics[width = 1.\columnwidth]{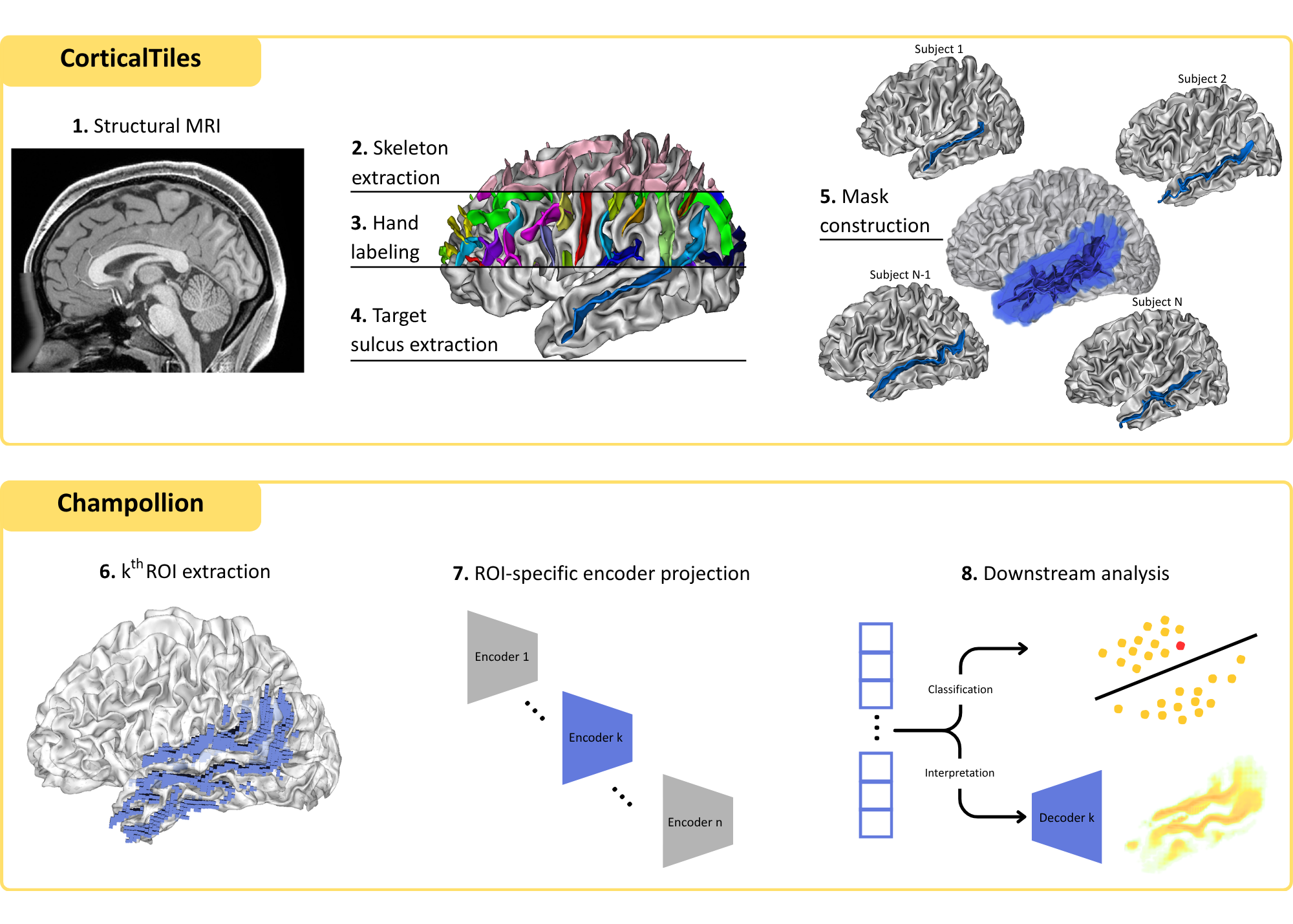}
\caption{\textbf{Pipeline overview}. \textbf{1.} Starting from structural MRI, \textbf{2.} BrainVisa/Morphologist extracts binary cortical skeletons following the medial surface of sulci, after affine registration to the MNI spatial referential~\cite{Evans1992-bd}. \textbf{3.} A small annotated dataset ($N=62$,~\cite{Perrot2011-pm,Borne2020-fp}) is used to construct a sulcus-specific mask in the MNI spatial referential (\textbf{4. and 5.}, illustrated here for the left superior temporal sulcus), \textbf{6.} enabling label-agnostic extraction of ROIs centered on the target sulcus (or set of neighboring sulci) for any subject. \textbf{7.} The resulting cortical skeleton tiles are encoded by a custom ROI-specific convolutional encoder ($\sim2$M parameters) to produce a latent representation of local folding patterns (32-dimensional representation space). \textbf{8.} These representations are used for downstream analyses. Each ROI is also associated with a dedicated decoder for interpretability.}
\label{fig:pipeline}
\end{figure}

\begin{figure}[H]
\centering

\begin{subfigure}[t]{0.41\textwidth}
    \centering
    \includegraphics[width=\linewidth]{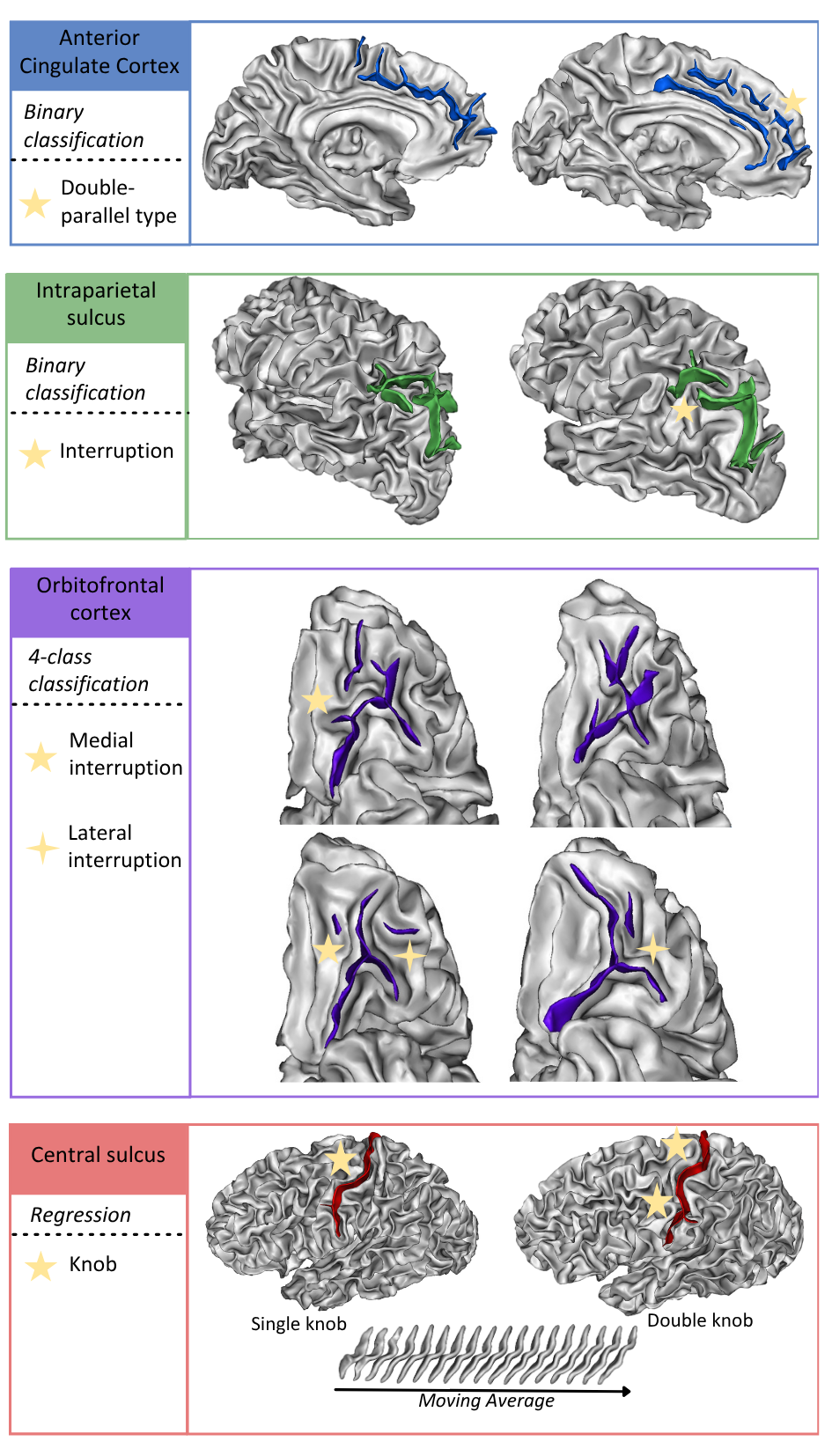}
    \caption{}
    \label{fig:patterns}
\end{subfigure}
\begin{subfigure}[t]{0.55\textwidth}
    \vspace{-300pt}
    \centering
    \vtop{
    \footnotesize
    \setlength{\tabcolsep}{4pt}
    \renewcommand{\arraystretch}{1.15}

\begin{tabularx}{\linewidth}{l *{4}{>{\centering\arraybackslash}X}}
    \toprule
    \textbf{Model}
    & \cellcolor{blue!50}\textbf{A. cingulate cortex}
    & \cellcolor{green!50}\textbf{Intra-parietal sulcus}
    & \cellcolor{purple!50}\textbf{Orbito-frontal cortex}
    & \cellcolor{red!50}\textbf{Central sulcus} \\
    & AUC & AUC & AUC & $R^2$ \\
    \midrule

    \multicolumn{5}{l}{\textbf{3D Binary Skeleton}} \\
    Champollion & \textbf{93.9} & \textbf{91.1} & \textbf{82.8} & \textbf{55.5} \\
    $\beta$-VAE~\cite{Guillon2024} & 73.0 & 66.9 & 60.2 & \underline{53.3} \\
    Baseline Linear & \underline{77.9} & 64.3 & 68.4 & 23.0 \\
    \addlinespace

    \multicolumn{5}{l}{\textbf{Zero-shot}} \\
    BSF~\cite{cox2024brainsegfounder} & 57.1 & 61.5 & 56.1 & 18.0 \\
    3DINO-ViT~\cite{xu2025threedino} & 69.6 & 73.4 & 68.5 & 43.3 \\
    SAM-Med3D~\cite{wang2025sammed3d} & 53.5 & 73.3 & 63.3 & 43.7 \\
    VISTA3D~\cite{he2025vista3d} & 45.7 & 71.9 & 75.5 & 45.7 \\
    DINOv3~\cite{simeoni2025dinov3} & 46.0 & 72.1 & 64.7 & 42.5 \\
    Point-M2AE~\cite{zhang2022pointm2ae} & 52.4 & 36.7 & 59.5 & 18.1 \\
    \addlinespace

    \multicolumn{5}{l}{\textbf{3D MRI Crop}} \\
    \multicolumn{5}{l}{\textbf{Zero-shot}} \\
    y-aware (VBM)~\cite{dufumier_exploring_2024} & -- & 47.4 & 55.3 & 0.0 \\
    BSF & -- & 62.4 & 54.9 & 2.1 \\
    3DINO-ViT & -- & 65.2 & 57.0 & 6.8 \\
    SAM-Med3D & -- & 64.2 & 54.1 & 18.9 \\
    DINOv3 & -- & 60.0 & 56.5 & 0.01 \\
    \addlinespace

    \multicolumn{5}{l}{\textbf{3D Binary Skeleton}} \\
    \multicolumn{5}{l}{\textbf{Self-supervised adaptation}} \\
    3DINO-ViT & 59.2 & \underline{75.8} & \underline{79.2} & 50.5 \\

    \bottomrule
\end{tabularx}

    \caption{}
    \label{tab:final_benchmark}
    }

\end{subfigure}

\caption{\textbf{Benchmarking of Champollion.} (a) The four downstream tasks used for benchmarking. Each task is performed on the ROI encompassing the relevant fold. For classification tasks, one representative sample per class is shown. For regression, six central sulcus descriptors are used, but only the first is illustrated. A moving average is displayed along with two samples at the extremes of the distribution. (b) Comparison between Champollion and baseline methods using linear probing on test sets not used for hyperparameter selection. For the orbitofrontal cortex task, which is a multiclass problem, we compute the one-vs-rest AUC and report the weighted average. For the regression task, we report the average $R^2$ score across the six descriptors. The baselines include a simple linear model using the flattened binary cortical tiles as input, a sulcal skeleton–specific self-supervised $\beta$-VAE retrained on the same dataset as Champollion~\cite{Guillon2024}, a self-supervised model trained on 10,000 whole-brain MRIs~\cite{dufumier_exploring_2024} after voxel-based morphometry~\cite{ashburner2000voxel} (VBM) preprocessing, and a range of computer vision foundation models, both medical~\cite{cox2024brainsegfounder,xu2025threedino,wang2025sammed3d,he2025vista3d} and natural image–based~\cite{simeoni2025dinov3,zhang2022pointm2ae}. For each task, best performance is in bold, second best is underlined. For the anterior cingulate cortex task, the MRI evaluation was skipped because we could not access the data.}
\label{fig:downstream_benchmark}

\end{figure}

The Champollion pipeline, from cortical folding extraction to local representation spaces, is illustrated in Figure~\ref{fig:pipeline}. Here, we define 56 overlapping ROIs to cover the whole brain. We adopt a lightweight convolutional architecture and the Barlow Twins objective~\cite{zbontar_barlow_2021} to incorporate domain-specific prior knowledge through carefully designed augmentations. Because only a limited number of folding patterns are well characterized, we benchmark Champollion using linear probing on four ROIs containing established patterns chosen to reflect the typical cortical folding variability. These tasks include (i) classification of sulcus interruption types (intraparietal sulcus, $N=390$, and orbitofrontal cortex, N=$577$), (ii) detection of sulcus presence or absence (anterior cingulate cortex, $N=381$), and (iii) regression of continuous shape descriptors (central sulcus, $N=880$) (Figure~\ref{fig:patterns}).

Champollion successfully captures all four tasks (Figure~\ref{tab:final_benchmark}). This performance primarily arises from the combination of tailored architectural and augmentation choices: (i) replacing the usual global pooling that follows the convolutional layers with a flatten-and-dense layer to preserve positional information in a setting where brains are globally aligned, (ii) applying mild positional augmentations to compensate for residual misalignment, and (iii) using a cutout-based masking strategy~\cite{devries_improved_2017} that maintains spatial consistency, in contrast to the standard crop-resize augmentation commonly used in natural image settings~\cite{chen_simple_2020,zbontar_barlow_2021,swav,byol}. Importantly, these components act synergistically, collectively outperforming a crop-resize-based pipeline on the four tasks (especially for the central sulcus regression task, $R²$ increases by $20.9\%$, Supplementary Table 6, 8). Additional gains are obtained by biasing the masking strategy to preserve sulcal fundi, thereby emphasizing folding features that are early-forming and stable over time ($7.6\%$ ROC-AUC increase on the orbitofrontal cortex task compared to standard cutout, Supplementary Figure 1).

By contrast, replacing the Barlow Twins objective with SimCLR~\cite{chen_simple_2020}, another augmentation-based method, yields comparable performance (Supplementary Table 11). Similarly, substituting the custom convolutional neural network ($\sim$2M parameters) with a 3D ResNet-18~\cite{resnet} architecture ($\sim$34M parameters) does not improve results despite increased computational cost (Supplementary Table 9). Optimal performance with ResNet-18 also requires replacing global pooling with a flatten-and-dense projection, indicating that the benefits obtained by preserving spatial information are not specific to the custom backbone.

\begin{figure}[h!]
\centering
\includegraphics[width = 0.85\columnwidth]{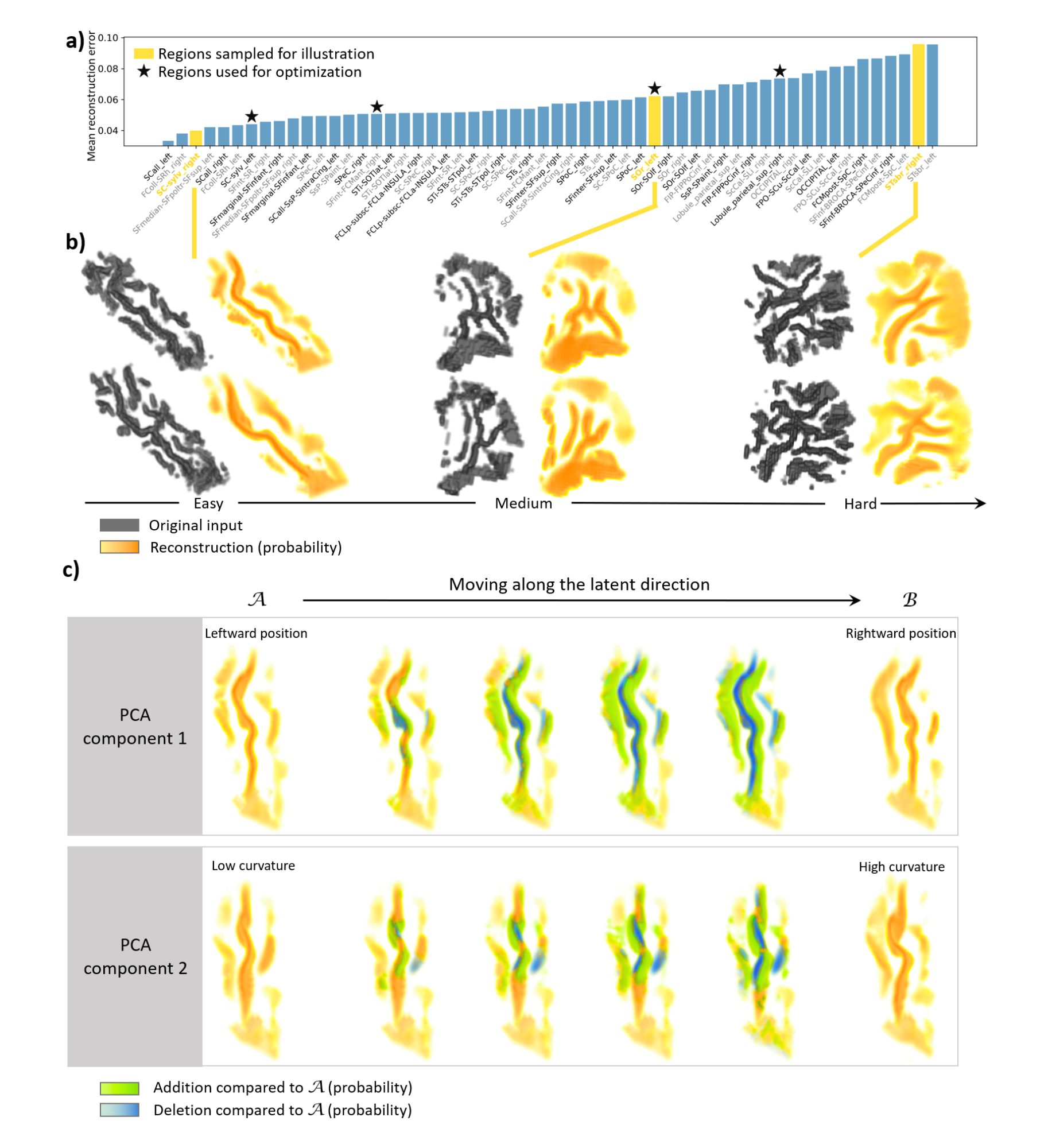}
\caption{\textbf{Decoder} trained on the UK Biobank dataset, independently of the encoder. \textbf{a)} Distribution of mean voxel-wise reconstruction error ratio on the UK Biobank test set across ROIs, ranked by error. Reconstruction quality is comparable across ROIs used for benchmarking and hyperparameter optimization (notably the augmentation strategy) and non-optimized ROIs, indicating that the training procedure generalizes beyond the four benchmarking tasks. \textbf{b)} Qualitative reconstructions on the HCP dataset (22-40 years, transfer from UK BioBank, 44-82 years) for three ROIs of increasing complexity (easy: right central sulcus; medium: left orbital sulcus; hard: ascending branches of the right superior temporal sulcus). Reconstructions are probability maps, with darker orange indicating higher predicted probability. \textbf{c)} Interpretation of latent directions through traversal. Main modes of variability of the right central sulcus are identified using principal component analysis (PCA), with the first two components shown. For each component, a subject from the ABCD dataset (9-11 years) is projected along the corresponding latent axis and decoded at successive quantiles of the component distribution (0.01, 0.25, 0.5, 0.75, 0.99). Reconstructions are shown as voxel-wise differences relative to the reference at quantile 0.01. The rightmost image corresponds to quantile 0.99. The first component primarily encodes a rostro–caudal translation of the sulcus, while the second captures variations in sulcal curvature. Along the translation component, as the sulcus shifts anteriorly, the postcentral sulcus progressively enters the ROI, maintaining a realistic gyrus width between the folds.}
\label{fig:decoder}
\end{figure}

To further assess the representation quality across all regions, we trained a simple decoder (inspired by Guillon et al.~\cite{Guillon2024}, Extended Data Figure~\ref{fig:decoder_backbone}). The main sulcal structures, typically located near the center of the ROIs, are reconstructed with high fidelity, even in complex regions (Figure~\ref{fig:decoder}). Because neighboring ROIs overlap, every fold is reconstructed centrally in at least one region. Therefore, Champollion provides a comprehensive description of the whole cortical skeleton.

We further provide a visual interpretation of the learned features through latent space traversal. As an illustration, we analyze the first two principal components of the representation space for the right central sulcus (Figure~\ref{fig:decoder}c). Traversal along the first component corresponds primarily to a translation of the sulcus along the rostro–caudal axis while preserving its shape. The second component reflects variations in sulcal curvature, with bending occurring around two anchor points. Overall, these traversals indicate that the representation captures semantically meaningful modes of folding variability.

\subsection{Relevant folding variability is missed by existing methods}

We compared Champollion with a set of representative baselines on the benchmark patterns, including a range of computer vision foundation models, both medical and natural image–based, in zero-shot evaluation (Figure~\ref{tab:final_benchmark}). For foundation models trained on MRI data, we tested both local cortical skeletons and ROI-centered MRI crops as input, and explored multiple representation extraction strategies (Methods).

Champollion consistently outperforms all zero-shot baselines across tasks. Importantly, foundation models fail to capture cortical folding information: their performances are poor when evaluated on local skeletons due to domain mismatch, and are even lower when applied to MRI crops, although in-domain, indicating that current neuroimaging foundation models largely overlook cortical folding information. We also note that a voxel-based morphometry preprocessing~\cite{ashburner2000voxel} (VBM) that introduces non-linear deformations in the input space completely erases folding information. Surprisingly, a naive linear baseline applied to the skeleton image space achieves moderate performance, suggesting that voxel-wise features retain predictive value after spatial normalization. We further assessed domain-specific models. A $\beta$-VAE trained on cortical folding skeletons~\cite{Guillon2024} is on par with Champollion for the central sulcus task, but does not capture the three classification tasks, that are more local and discrete. As for foundation models, we selected the best performing model in the zero-shot setting, namely 3DINO-VIT~\cite{xu2025threedino}, and performed self-supervised adaptation (continual pretraining). Adaptation narrows the gap with Champollion, but still yields lower performance, despite substantial computational cost.

We next quantified the correspondence between Champollion’s representations and classical sulcal descriptors extracted with the BrainVisa Morphologist pipeline (computed for each sulcus of an atlas in the MNI space, Methods, detailed results in Extended Data, Figure~\ref{fig:morpho}).
While morphometric descriptors directly related to sulcal shape are partially reflected in Champollion’s representations, low correlations in some ROIs do not reflect a loss of relevant information. Indeed, by applying a multivariate genome-wide association framework on both sets of phenotypes (Methods) in one ROI with weak-to-moderate correlations (right calcarine and intralingual sulci, ScCal–SLi) and one ROI with strong correlations (left inferior frontal and inferior precentral sulci and Broca's area, SFinf–BROCA–SPeCinf), Champollion encompasses and exceeds the genetic associations obtained with classical morphometry in both cases.
In left SFinf–BROCA–SPeCinf, Champollion identifies 43 independent genetic loci, compared with 10 loci detected using classical morphometric descriptors. Of these, 9 loci overlap between the two approaches, with identical lead SNPs for 7 loci. Similarly, in right ScCal–SLi, Champollion identifies 122 loci, whereas classical morphometry identifies only 9, 8 of which overlap with Champollion’s results, including 3 shared lead SNPs (Extended Data, Figure~\ref{fig:miami_plot}).
Overall, these results support the novelty of Champollion's phenotype.

\subsection{Main variability modes are largely independent across sulci}

\begin{figure}[h!]
\centering
\includegraphics[width = 0.7\columnwidth]{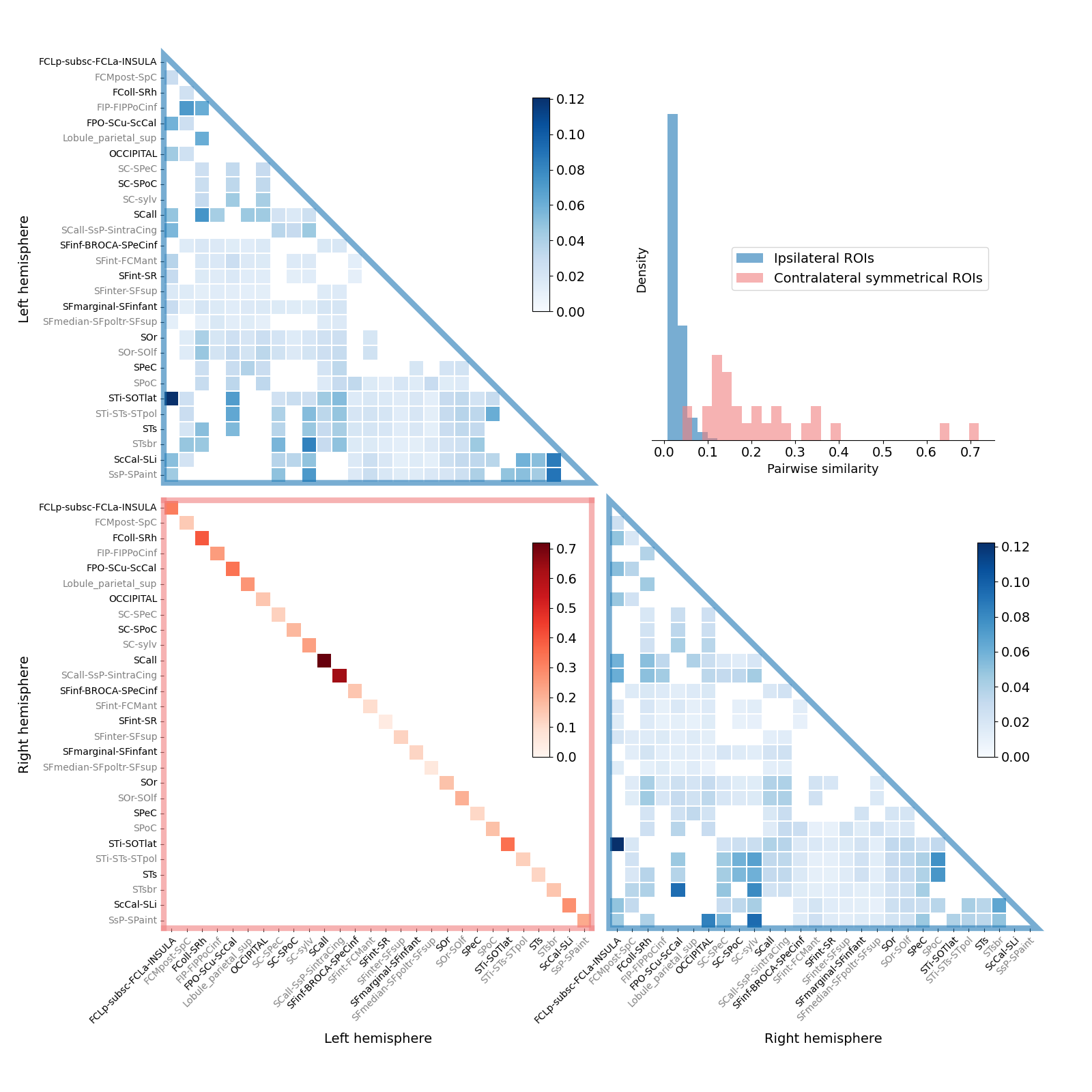}
\caption{\textbf{Pairwise similarity between regional representation spaces}, computed using linear centered kernel alignment (CKA) on the HCP dataset. Overlapping regions (Dice score $>10\%$) and contralateral non-symmetric regions are excluded. Top right: distribution of pairwise similarities. Top left and bottom right: pairwise similarities between ipsilateral regions. Bottom left: pairwise similarities between contralateral homologous regions. Symmetric ROIs exhibit substantially higher similarity than non-symmetric, non-overlapping regions. Of note, contralateral non-symmetrical ROIs were masked because of possible confounding factors such as residual similarity between symmetrical regions propagating to overlapping regions. For instance, the similarity between the left pre-central ROI and the right post-central ROI could arise from the overlap of both regions with their respective ipsilateral central regions that have genuine similarity. Region names correspondence to sulci is given in Extended Data, Table~\ref{tab:acronyms}.}
\label{fig:similarity_main}
\end{figure}

We quantified the similarity between the regional representations obtained with Champollion using linear Centered Kernel Alignment (CKA)~\cite{cka}, a standard method for assessing representational similarity in deep learning models. Across all non-overlapping ROIs within the same hemisphere (dice score $<10\%$), we observed systematically low pairwise similarity ($< 12\%$; Figure~\ref{fig:similarity_main}), indicating that the sulcal regions encode largely independent shape information. This finding is consistent with previous morphometric analyses~\cite{Snyder2024}, which reported a median absolute cross-correlation of only $3\%$ between 200 sulcal phenotypes derived from 5 morphometric measurements across 40 sulci. However, we report an exception for contralateral homologous regions which exhibit higher similarity.

We further divided the training dataset into two halves, and for each split, we randomly selected five ROIs to train separate models. Representations obtained for the same region show high similarity ($95-98\%$), demonstrating both training stability and robustness to dataset sampling.

Overall, the consistently low similarities reflect largely independent main modes of variability across ROIs, supporting our strategy of learning folding representations locally.

\subsection{Champollion is biologically relevant on a range of datasets}

We further assessed whether Champollion relates to two biologically grounded tasks: sex prediction and twins pairing. Accordingly with the previous section suggesting a complementarity of the ROIs, we iteratively aggregated the latent variables from the regional spaces to increase signal, eventually building a whole-brain representation. 

\begin{figure}[t]
    \begin{subfigure}[t]{0.48\textwidth}
    \vspace{-170pt}
        \begin{subfigure}[t]{\linewidth}
            \includegraphics[width=\linewidth]{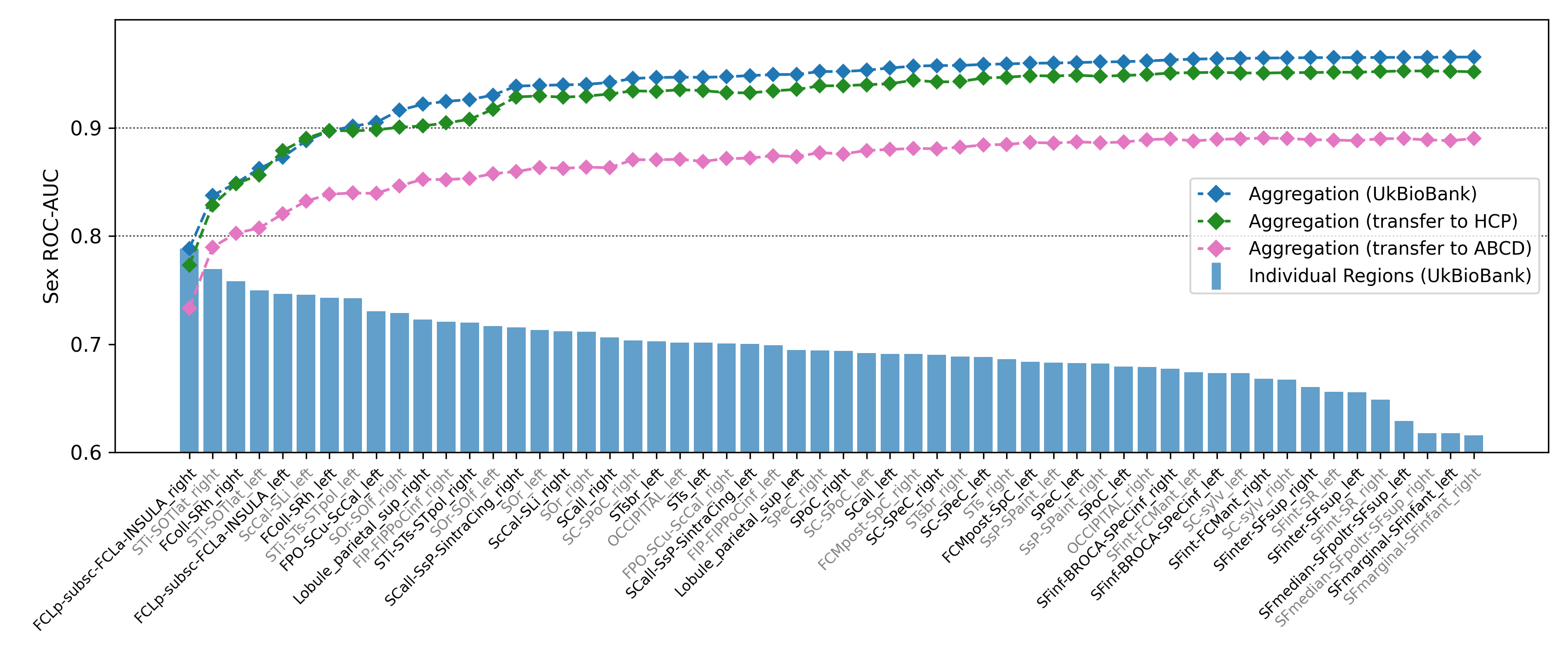}
            \caption{}
            \label{fig:sex_classification}
        \end{subfigure}
        
        \begin{subfigure}[t]{\linewidth}
            \includegraphics[width=\linewidth]{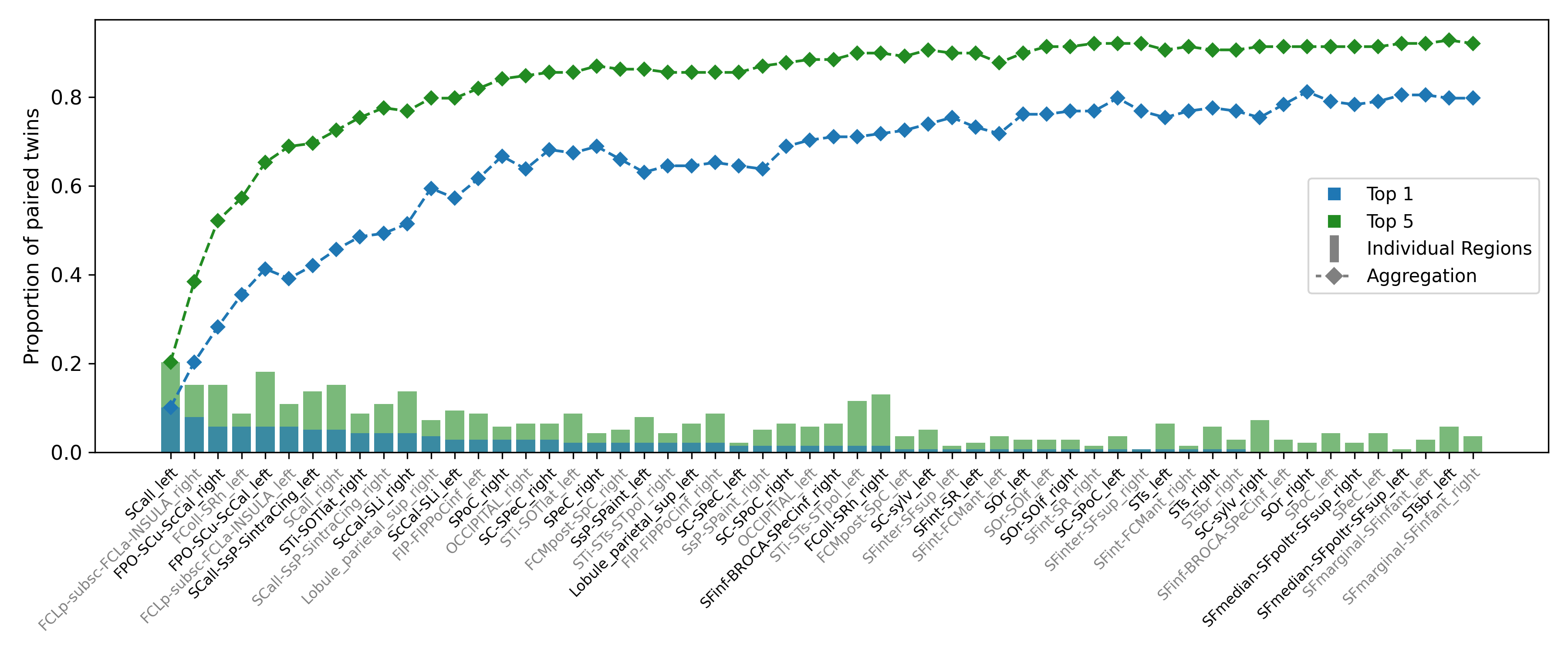}
            \caption{}
            \label{fig:twins_identification}
        \end{subfigure}
    \end{subfigure}
    \hfill
    \centering
    \begin{subfigure}[t]{0.48\textwidth}
        \includegraphics[width=\linewidth, trim={0 0 0 1.5cm}, clip]{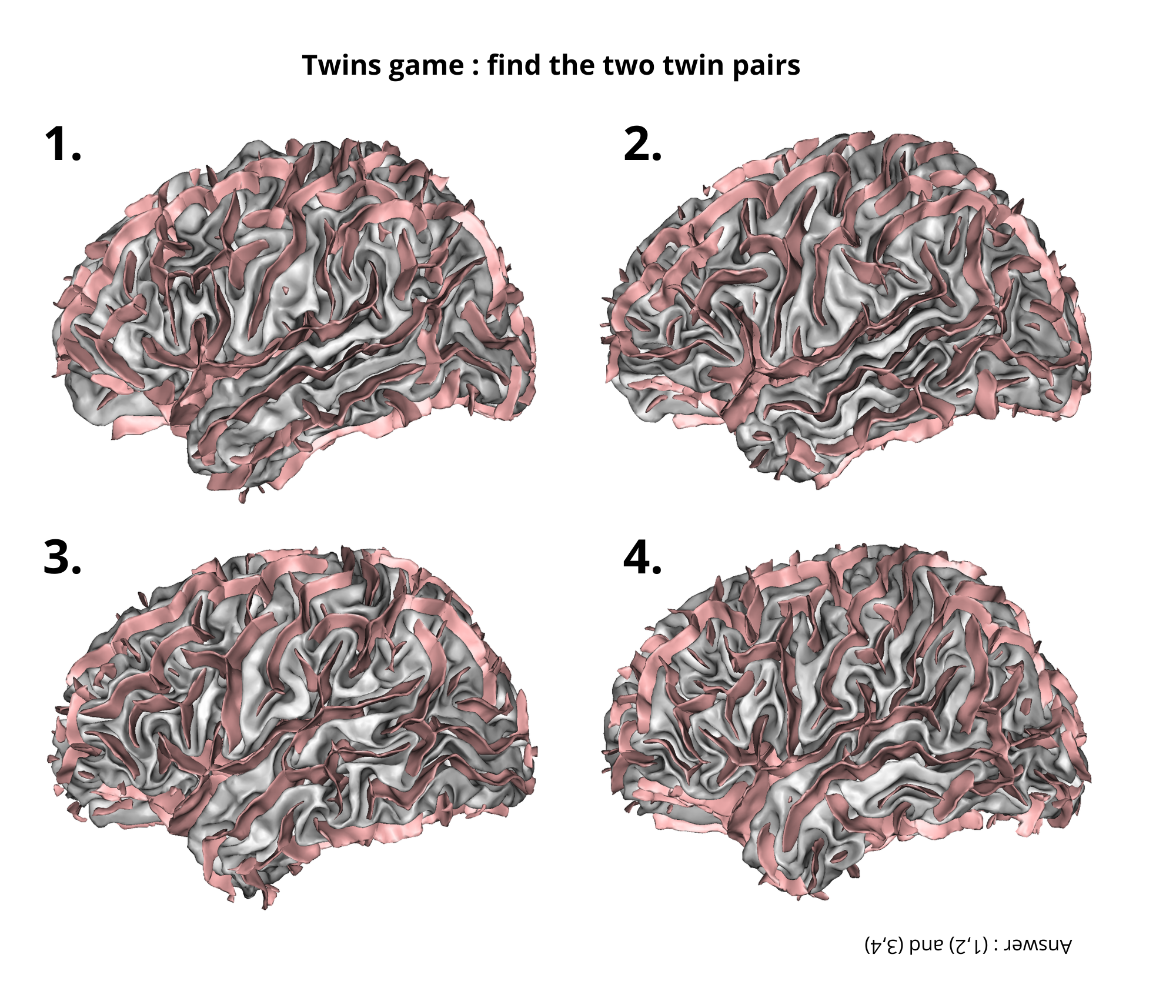}
        \caption{}
        \label{fig:twins_game}
    \end{subfigure}
    \caption{\textbf{Champollion captures biologically meaningful information.} \textbf{a)} Sex classification from regional cortical representations across the whole brain. Regions are iteratively aggregated by representation space concatenation, ordered by decreasing individual performance. Additional curves report direct transfer of the classifier hyperplane obtained on UK Biobank to external datasets with varying domain gaps, from younger adults (HCP, $<1\%$ AUC loss) to children (ABCD, $7\%$ AUC loss). \textbf{b)} Monozygotic twin identification using nearest-neighbor pairing in the HCP dataset (Euclidean distance in the representation space). Region names correspondence to sulci is given in Extended Data, Table~\ref{tab:acronyms}. \textbf{c)} The twins game: the four brains correspond to two monozygotic twin pairs, aligned in the MNI spatial referential. We invite the reader to recover the pairs. The answer is written upside down. This game illustrates how variable cortical folding is, as even monozygotic twins exhibit substantially different patterns.}
\end{figure}

Doing so, sex is predicted with $97\%$ AUC (Figure~\ref{fig:sex_classification}), which is comparable to what is typically achieved using all the structural MRI information~\cite{dufumier_exploring_2024}. Moreover, while representations obtained from MRI suffer from site effects and show poor transferability to external datasets, the classification hyperplane can here be transferred with marginal loss of performance from UK Biobank to the younger adult dataset HCP ($<1\%$), despite a domain gap induced by age, acquisition site and scanner resolution (Extended Data Table~\ref{tab:datasets}), owing to the preprocessing pipeline and the time stability of the patterns.

Monozygotic twins are paired on the HCP dataset (138 pairs among 1114 subjects, Figure~\ref{fig:twins_identification}), based on Euclidean distance in the representation space. The best performing individual ROIs correspond to the first forming sulci (callosal sulcus, sylvian fissure, parieto-occipital fissure, ~\cite{chi1977gyral}), consistent with previous studies suggesting that the earliest forming sulci are the most genetically regulated~\cite{Pizzagalli2020-qy}. When concatenating all regional representations, $80\%$ are nearest neighbors and $92\%$ are in the top 5.

Following these experiments, we provide further evidence that representations are biologically meaningful, and we demonstrate that they naturally transfer to external datasets. Next, the model is used to perform clinically-relevant tasks on various datasets.

\subsection{Brain-wide discovery and interpretation of localized folding signatures}

\begin{figure}[H]
\centering
\includegraphics[width = 0.95\columnwidth]{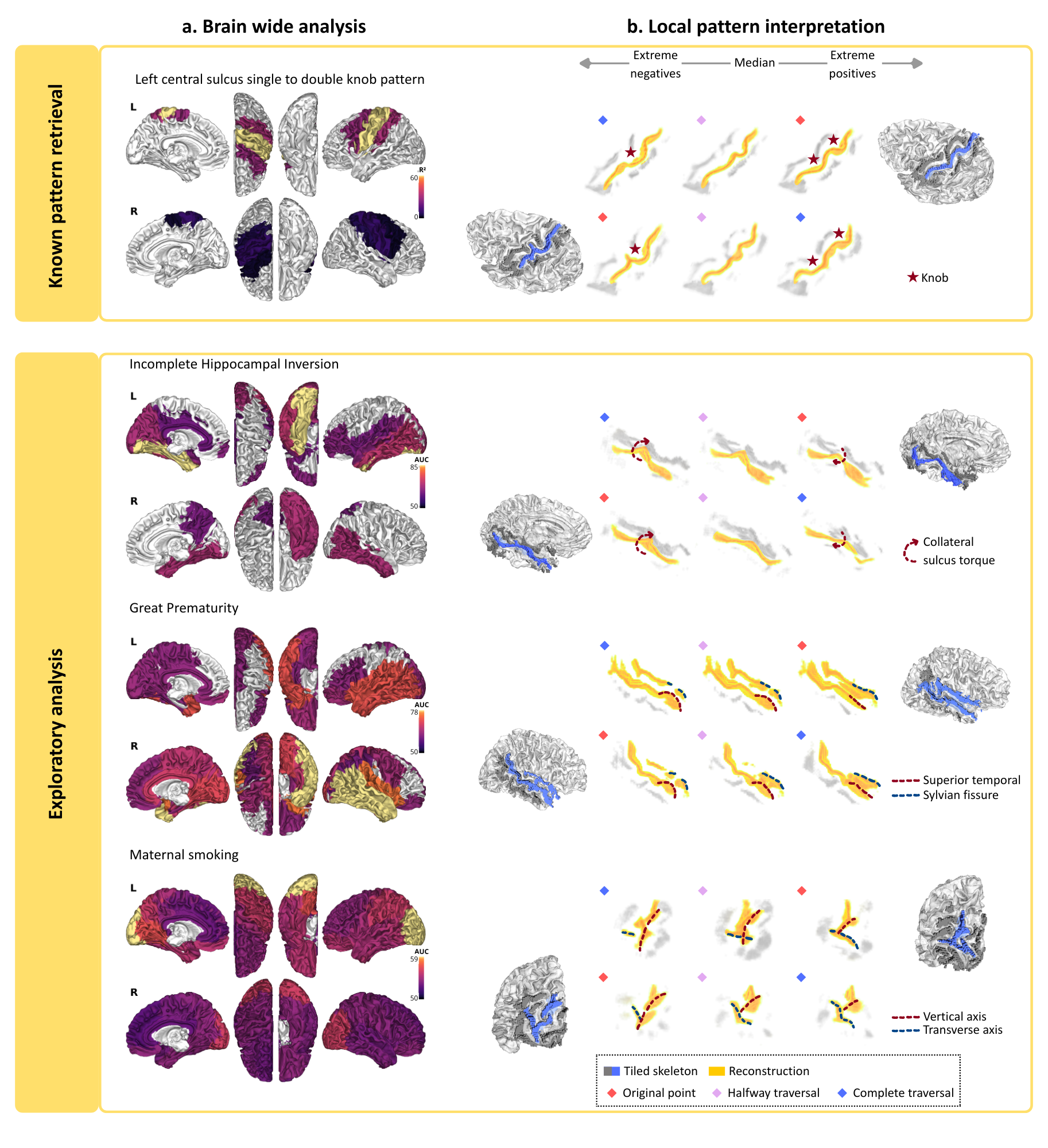}
\caption{\textbf{Regional associations between cortical folding and phenotypes.}
\textbf{Left:} brain-wide maps of classification performance, with non-significant ROIs shown in gray (Methods). \textbf{Right:} visualization of discriminative folding patterns for the top-performing ROI in each task. For each task, latent traversal is performed along the classification direction starting from well-classified positive and negative subjects, and decoded using the Champollion decoder (Figure~\ref{fig:decoder}). Reconstructions are shown at the extremes and median of the distribution (quantiles 0.01, 0.5, 0.99). Due to space constraints, two subjects are displayed per task; additional examples are provided in Supplementary Figure 15-19 to better interpret the consistent variations across subjects. Sulci not involved in the illustrated patterns are shown in gray for clarity. \textbf{Top:} validation task using the central sulcus single- to double-knob pattern~\cite{sun_effect_2012}, showing a localized signal and a consistent transition. \textbf{Second:} incomplete hippocampal inversion, with peak performance in the left collateral sulcus (AUC = 0.85) and additional signal in bilateral temporal and calcarine regions, revealing a torque signature of the collateral sulcus. \textbf{Third:} great prematurity, with dominant signal in the right superior temporal cortex (AUC = 0.78), interpreted as a subtle knob flattening of the anterior part of the superior temporal sulcus. \textbf{Bottom:} maternal smoking exposure during pregnancy, which produces weaker but significant associations in the UK Biobank (AUC = 0.59), primarily within occipital regions. In the left occipital lobe, the signal corresponds to a global reorientation of the main folds.}
\label{fig:exploratory_analysis}
\end{figure}

We evaluated whether Champollion can identify localized cortical folding signatures associated with diverse biological phenotypes. Using a unified analysis framework (Methods), we first generated brain-wide maps of regional predictive performance by linear probing of the learned representations, then interpreted the most informative regions through latent traversal (Figure~\ref{fig:exploratory_analysis}). We considered four increasingly challenging applications, spanning a known cortical folding pattern, a developmental anatomical variant, an early-life condition, and an environmental exposure.

We first validated the framework on the well-characterized single- versus double-knob configuration of the left central sulcus, a label obtained with Isomap and previously used as a downstream task (Figure~\ref{fig:patterns}, HCP, $N=880$). Consistent with the low observed correlations between the folds, predictive performance is almost exclusively localized to the corresponding cortical region, and latent traversal faithfully recovers the reference folding pattern. Although regression on the Isomap coordinate remains imperfect ($R^2=60\%$), a substantial fraction of the remaining error likely reflects Isomap label noise rather than model limitations, since the latent traversal closely matches a denoised version of the Isomap trajectory (obtained with a moving average), suggesting that Champollion encodes smoother variations than Isomap.

We then investigated three neurodevelopmentally relevant phenotypes: incomplete hippocampal inversion (left hemisphere, QTIM, 23±3 years, $N=928$, $8\%$ positives), great prematurity (ABCD, 9-11 years, $N=8{,}846$, $1\%$ positives), and maternal smoking during pregnancy (56±8 years, $N=37{,}925$, $30\%$ positives). Across all applications, predictive information remains spatially confined to a limited number of cortical regions, while latent traversal consistently yields coherent and interpretable shape variations. The recovered directions are highly stable under bootstrap resampling, and even the least stable task (prematurity) produces visually consistent interpretations across bootstrap iterations (Supplementary Figure 17-18). Remarkably, the maternal smoking signature identified in the left occipital lobe in UK BioBank (56±8 years) is recovered in the independent ABCD dataset (9-11 years) through a direct hyperplane transfer from the UK BioBank fit, despite the large age difference between cohorts. Moreover, the ABCD prediction score increases with increasing tobacco exposure (Extended Data Figure~\ref{fig:abcd_smoking_auc}). Together, these results demonstrate that Champollion provides a unified framework for discovering, localizing and interpreting long lasting cortical folding signatures.

\section{Discussion}

We introduce Champollion, a self-supervised framework for learning local, interpretable representations of cortical folding from structural MRI. Across multiple cortical regions and datasets, Champollion consistently captured folding patterns better than all tested foundation models, showing that cortical folding remains largely underrepresented in current neuroimaging representations.

The local formulation of Champollion revealed a high degree of independence between folding patterns across cortical regions in the general population. Thus, in addition to bringing native interpretability, such design is appropriate to capture the whole brain variability. In particular, this cortical folding peculiarity could partly explain the failure of foundation models: such localized information may be overlooked by neuroimaging models  that are optimized to capture broad anatomical variability.

We further highlighted how the oversight of cortical folding in generic neuroimaging representations constitutes a missed opportunity, using Champollion to showcase the richness and interpretablility of the information embedded in folding patterns. In particular, folding patterns are genetically grounded: the ability to match monozygotic twins based on latent proximity provides evidence that the learned features reflect heritable aspects of cortical folding, and Champollion's representations exhibit substantially stronger genetic associations than the classical morphometric descriptors of folding.

We further evaluated the practical utility of Champollion through exploratory analyses on several tasks of increasing difficulty. A simple latent traversal framework identified subtle but consistent shape changes across samples in the most informative regions. Here, it included a torque reduction of the left collateral sulcus for incomplete hippocampal inversion (IHI), a flattening of the anterior right superior temporal sulcus for prematurity, and a global reorientation of folds in the left occipital region for maternal smoking.

Notably, the maternal smoking pattern was replicated across the UK Biobank (44–82 years) and ABCD (9–11 years) cohorts, suggesting that the detected signal reflects a stable, long-term signature. Together with the successful transfer of classifiers across datasets with markedly different demographics, this result indicates that Champollion is robust to site and population variability. The identification of a coherent pattern in the occipital region is particularly notable, as this region is highly variable and lacks a standardized sulcal nomenclature in anatomical literature.

Here, we relate the patterns identified in the exploratory analysis to the existing literature. The association between IHI and the ipsilateral collateral sulcus is anatomically consistent with the location of the hippocampus. Notably, de Matos et al.~\cite{de_matos_temporo-basal_2023} reported no association between left IHI and manually defined sulcal variants in the collateral region, primarily based on interruption patterns. This aligns with our findings, as the identified IHI-related signature is a previously unreported torsion of the collateral sulcus. For prematurity, the temporal lobe is consistently implicated in prior studies~\cite{Benjamin2025-zp,Mancuso2025-gl}. Premature birth disrupts exposure to the intrauterine auditory environment, which may affect the maturation of auditory regions located in the temporal lobe. In line with this, Benjamin et al.~\cite{Benjamin2025-zp} showed that STS depth at term-equivalent age correlates with gestational age at birth, with stronger effects in the right hemisphere, and that auditory stimulation (music exposure) modulates this effect. While the study focuses on sulcus depth, our results suggest that prematurity may also be associated with more subtle alterations in folding patterns, extending beyond classical morphometric measures. In contrast, for maternal smoking, we did not identify prior reports of associations with sulcal patterns in the occipital region. Existing studies mainly report weak effects on cortical thickness~\cite{El_Marroun2014-bv,derauf_subcortical_2012}. The folding pattern identified here may therefore reflect previously uncharacterized structural alterations. We however emphasize that these analyses illustrate a general framework for hypothesis generation rather than establishing definitive biological associations.

More broadly, while recent neuroimaging models aim to identify biomarkers directly from structural MRI~\cite{dufumier_exploring_2024,barbano2025anatomical}, such approaches often conflate developmental markers with long term disease consequences. In contrast, cortical folding patterns emerge early and remain stable over time, making them promising candidates for probing neurodevelopmental mechanisms and causal pathways.

From a practical perspective, Champollion is well suited for clinical use. It relies on standard and widely available structural MRI, requires only moderate spatial resolution (2\,mm isotropic), and produces compact, region-specific representations that can be analyzed with simple statistical models.

This work has several limitations. First, the ROI atlas is one possible tiling rather than a unique optimal decomposition. Second, the identified patterns may correlate with other anatomical factors such as local brain volumes and global shape. Although prior work suggests that these factors are not the primary determinants of folding variability~\cite{troiani_variability_2022}, further analysis is required to determine their specific contribution to the patterns identified here. Third, combining local and global representations may reveal coordinated patterns lying in low-variance modes. Looking forward, we will derive an anomaly detection framework from this model of typical variability, with potential applications to developmental disorders such as epilepsy~\cite{Regis2011-kq}, psychiatry~\cite{Cachia2021-vb} and rare genetic syndromes~\cite{Snyder2026}.

\section{Methods}

We sought to learn self-supervised representations of cortical folding variability based on structural MRI. We first provide details on the pipeline, then on how Champollion was optimized, and finally on the downstream analyses performed in the study.

\subsection{Preprocessing}

Brain MRI data present several challenges for representation learning: image volumes are large and semantically dense relative to the number of available samples, inter-individual variability is high, and acquisition biases such as age effects or site differences are prominent. Standard pipelines such as voxel-based morphometry~\cite{ashburner2000voxel} or surface-based registration~\cite{Fischl2012-fy} address part of this variability through non-linear alignment, but such approaches inevitably distort the cortical folds and therefore discard precisely the information we seek to capture. Our goal was to develop a preprocessing strategy that isolates folding shape without relying on non-linear normalization, thereby reducing acquisition biases while preserving the sulcal patterns. To this end, we designed a three-step preprocessing pipeline based on cortical skeletonization, global affine alignment of the brains on a template, and region-wise cropping.

\textbf{Cortical skeletonization.} Sulcal folds, once fully developed in early childhood, have a stable shape and topology throughout life, whereas aging primarily affects folds opening. We therefore extract the folds as one-voxel-wide skeletons that trace the medial axis of the cortical surface perpendicularly to the brain hull to remove the opening information. Practically, we process structural MRI volumes using the BrainVisa Morphologist pipeline, which produces a skeletonized negative cast of the white matter surface. This procedure converts sulci into 3D surface-like objects of one-voxel thickness, faithfully following the center of each fold, using a watershed-based algorithm relative to image intensities~\cite{mangin_tmi}.

\textbf{Affine alignment.} The resulting cortical skeletons are then affinely aligned to a common template, the MNI spatial referential (ensuring that the brains have the same orientation and size), and resampled at a 2\,mm isotropic resolution, with the skeleton constrained to remain one voxel wide.

\textbf{Regional tiling.} While the main folds are present in all individuals, its their shape rather than their presence that vary substantially. The spatial normalization thus allows us to divide the cortical skeleton into regions that anatomically correspond across subjects. We tile the cortical skeleton into ROIs defined in the MNI spatial referential to encompass one or a few neighboring sulci of interest, and learn local representations based on these tiles rather than a global representation based on the whole cortical skeleton, for two main reasons: (i) such design aligns with established practices in the folding community, where analyses are typically conducted on predefined regions of interest, (ii) it provides inherent interpretability, with latent variables directly tied to anatomical regions and enabling spatial maps of association as well as visual characterization of individual sulcus folding patterns. To define the ROIs, we used an independent dataset~\cite{Perrot2011-pm} ($N=62$) in which all sulci were hand-labeled according to the BrainVisa atlas. For each ROI, we take the union of the corresponding atlas-defined sulci across all 62 subjects and apply a 10\,mm isotropic dilation, ensuring that the tile encompasses inter-individual and inter-dataset variability. This produces robust overlapping ROI masks that can be consistently applied to any other dataset, using the spatial normalization to the MNI referential. The full preprocessing pipeline is summarized in Figure~\ref{fig:pipeline}. Here we divided the cortical skeleton into 56 overlapping ROIs, corresponding to 28 ROIs per hemisphere, using our CorticalTiles toolbox. Importantly, we treated the left and right hemispheric ROIs separately, since the sulcal regions are not guaranteed to be bilaterally symmetric.

\textbf{Choice of the sulcal regions.} We define 56 overlapping ROIs to cover the whole brain, each encompassing one or a few sulci of interest. The ROI definition remains flexible and may consist of any combination of neighboring sulci in the atlas, provided the resulting patch is of manageable size for the lightweight backbone. The sizes of all 56 ROIs used in this study are reported in the Supplementary Figure 14 and provide an estimate of the acceptable range. The correspondence between the sulcus names and their acronyms is provided in Extended Data, Table~\ref{tab:acronyms}, and a visualization of the ROIs in the MNI spatial referential is provided in Supplementary Figure 12-13.

\subsection{Self-supervised learning}

Most SSL approaches in computer vision fall into two broad categories~\cite{assel2025jointembedding}: reconstruction-based models, which learn to decode the input from a compressed representation~\cite{higgins2017betavae,He2022-yv}, and joint-embedding models, which enforce consistency between augmented views of the same sample~\cite{chen_simple_2020,zbontar_barlow_2021,byol,chen2020simsiam,Caron_2021_ICCV,JEPA}. The latter paradigm enables the explicit injection of prior knowledge through carefully designed augmentations. In the context of cortical folding, these include tolerance to small positional variations after global affine alignment to the MNI spatial referential, and a focus on sulcal fundi that constitute the earliest traces of folding dynamics. Since we want to use lightweight local encoders, we choose a convolutional architecture, and therefore exclude the SSL methods originally designed for transformers (MAE~\cite{He2022-yv}, JEPA~\cite{JEPA}, Ibot~\cite{zhou2021ibot}, and Dino~\cite{Caron_2021_ICCV}). Since we want to use custom augmentations that go beyond masking, it further excludes MAE and JEPA. Because SimCLR~\cite{chen_simple_2020} relies on negative pairs, which are ill-defined in our context without clear class boundaries, we adopt the Barlow Twins loss~\cite{zbontar_barlow_2021}. It aligns the representations of paired augmentations while simultaneously penalizing correlations across features, thereby encouraging each dimension to carry non-redundant information. Additionally, BYOL~\cite{byol} and SimSiam~\cite{chen2020simsiam} rely on implicit mechanisms to prevent collapse and can exhibit dimensional collapse under unfavorable optimization regimes~\cite{SimSiamCollapse,liu2022bridging}, whereas Barlow Twins explicitly regularizes feature covariance and is theoretically less prone to collapse. We set the redundancy reduction parameter to $\lambda = 8\times10^{-2}$.

\subsection{Data augmentations for SSL}

The choice of augmentations encodes prior knowledge of the data and can strongly influence the nature of the encoded features.

\textbf{From natural images to cortical skeletons.} For natural images, there is broad consensus on useful augmentations. Typical natural imaging pipelines~\cite{chen_simple_2020} employ Gaussian noise or blur (to mimic resolution variability), horizontal flips (perspective invariance), color jittering and grayscale conversion (robustness to hue and texture), and most importantly, cropping. The most widely adopted cropping strategy is crop-resize, that fosters scale invariance and the learning of statistical dependencies across image parts. Cropping is therefore regarded as an essential augmentation for building robust, abstract representations across vision domains~\cite{balestriero_cookbook_2023}.

In medical imaging, however, there is no established consensus on augmentation policies~\cite{dufumier_exploring_2024}. Cortical skeletons present additional challenges: they are binary, textureless images aligned to a common orientation, rendering blur, color jittering, and flipping irrelevant. Cropping remains the only natural inspiration, but it is here adapted to respect spatial normalization.

\textbf{Interplay of augmentations and backbone design.} Because the skeletons are globally aligned to the MNI spatial referential, local shape features should not be disentangled from their spatial location. For example, a knob or interruption in a fold carries distinct meaning depending on whether it lies in an anterior or posterior segment. Standard CNN backbones such as ResNet, which rely on translation equivariance with global pooling, seem therefore suboptimal. We instead design a lightweight 3D CNN whose final pooling layer is replaced by a flatten operation, followed by a dense projection layer. This architecture preserves positional information in the representation space while retaining the inductive biases of convolutions.

At the same time, affine global spatial normalization is not providing perfect alignment across brains at the local scale: inter-individual differences in brain shape induce slight local misalignments. To avoid overfitting to absolute position, we introduce mild translation and small rotations in the augmentation pool. These augmentations exploit the Barlow Twins loss to enforce local positional invariances and thus complement the positional sensitivity of the backbone. In practice, some positional features are still learned (Figure~\ref{fig:decoder}), but the shapes are much better learned when adding these positional regularizations.

\textbf{Masking strategies: cut-in and cut-out.} We implement cut-out~\cite{devries_improved_2017}, which masks voxels inside a random bounding box placed uniformly across the ROI tile. It replaces the usual cropping augmentation. Unlike crop-resize, the images are not re-centered and rescaled, ensuring consistency of spatial positions. To complement cut-out, we implement cut-in. High variability arises at ROI boundaries, where neighboring folds may appear or disappear across individuals depending on misalignment. To regularize this effect, cut-in masks voxels outside a random bounding box centered on non-zero voxels. Cut-in acts as an edge regularizer: it biases masking toward peripheral regions, which may remove edge information but leaves central fold shape intact. Since the cortical ROIs overlap, no information is permanently excluded at the whole-brain level.

\textbf{Incorporating a focus on sulcal fundi.} Each skeleton is decomposed into simple surfaces, which are topologically elementary surfaces without inner junction, and fundus voxels are segmented~\cite{Mangin1995-qp}. Cut-in and cut-out are modified to selectively preserve the fundus voxels with a tunable probability, sampled at the simple surface level. Preserving some fundus information encourages the network to focus on the most semantically meaningful voxels, as they constitute the earliest traces of folding dynamics. Importantly, always preserving bottom voxels trivializes the task and yields poor representations (Supplementary Figure 1), thus the stochastic sampling.

\textbf{Complete augmentation policy.} The final augmentation pipeline applies (i) random rotations (up to 18° per axis, uniform sampling) and translations (±1 voxel per axis, uniform sampling) on each axis with probability 1 (since the sampling is centered on zero), followed by (ii) masking (cut-in or cut-out) with probability 0.8. Cut-in and cut-out are applied exclusively, with parallelepiped bounding boxes of same proportion as the ROI, their size being expressed as a volume ratio, thereby independent of the ROI. For cut-in, we sample uniformly in the range 20\%-90\% of the volume (corresponding to 58.5\%-96.5\% per axis). As for cut-out, we sample in the range 5\%-30\% of the volume (36.8\%-66.8\% per dimension). For both cut-in and cut-out, the fundus voxels are preserved with probability 0.5. This unified hyperparameterization is necessary for whole-brain modeling, where 56 distinct ROIs are processed with a common policy. A detailed ablation study on these augmentations and their hyperparameters is provided in Supplementary section 1.

\subsection{Backbone}

The backbone is a custom 12-layer 3D CNN with approximately 2 million convolutional parameters. It employs kernels inspired by ResNet ($7\times7\times7$ in the first layer, $3\times3\times3$ thereafter) and applies three downsamplings (stride 2 at layers 4, 8, and 12, along with an increasing number of filters: 1, 32, 64, 128). Because the ROIs are relatively small, deeper or wider networks were not required. Contrarily to ResNet, no residual connections were required due to the shallow architecture.

Following the convolutional layers, the feature maps are flattened ($\sim8,000 - 40,000$ features depending on ROI size) and projected through a dense layer into a 32-dimensional representation space. This design eliminates translation equivariance, thereby binding local features to their spatial locations. To optimize training under the Barlow Twins loss, we append a three-layer non-linear MLP projection head, which expands the representations to 128 dimensions for the loss computation (as recommended in~\cite{zbontar_barlow_2021}).

\begin{figure}[p]
\centering
\includegraphics[width = 1.\columnwidth]{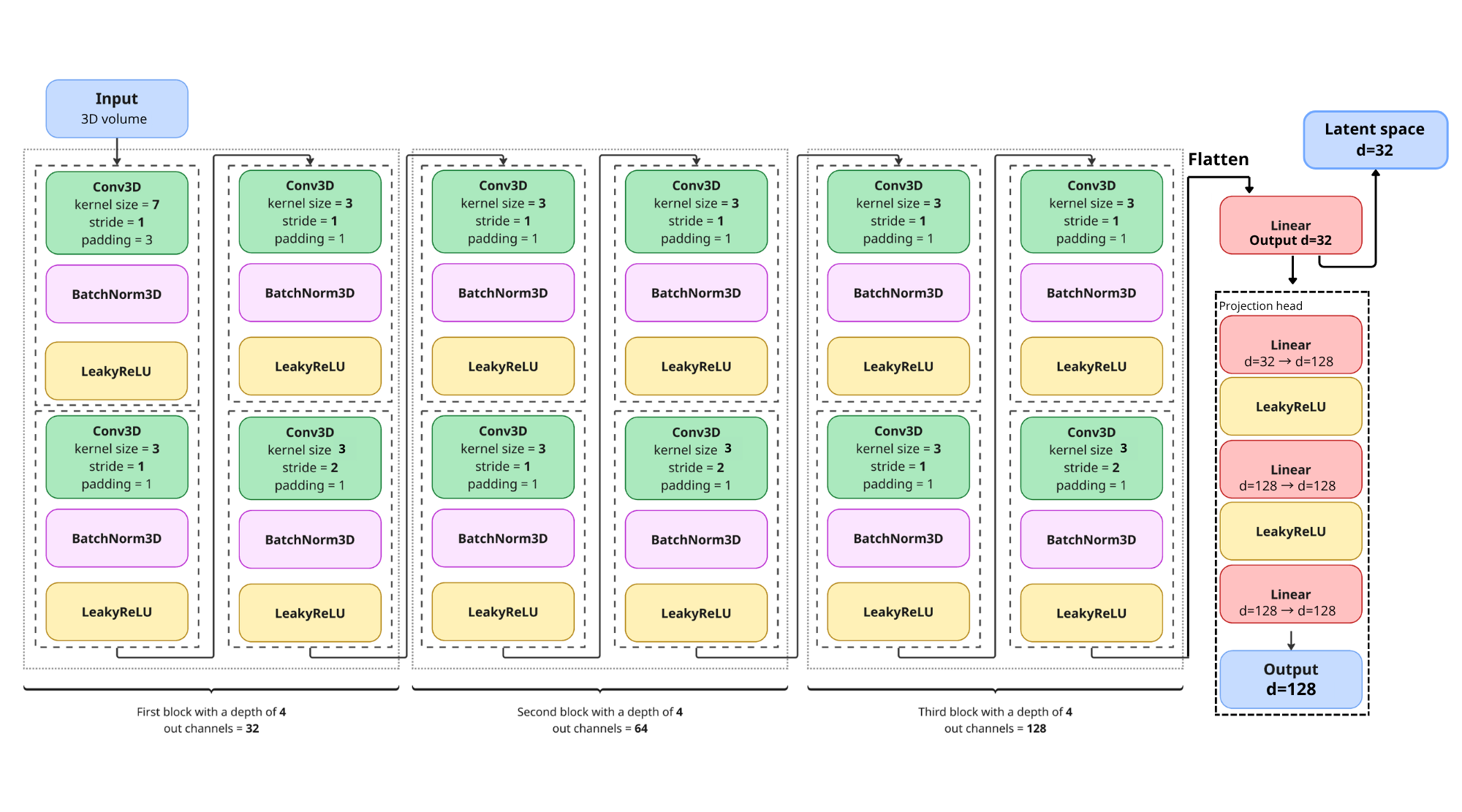}
\caption{Champollion's backbone: a 12-layer convolutional neural network that embed local crops of cortical skeletons in a 32-dimensional representation space. As in standard practice with the Barlow Twins loss, a non-linear projection head is added between the representation space and the output space (in which the loss is computed).}
\label{fig:backbone}
\end{figure}

\begin{figure}[p]
\centering
\includegraphics[width = 1.\columnwidth]{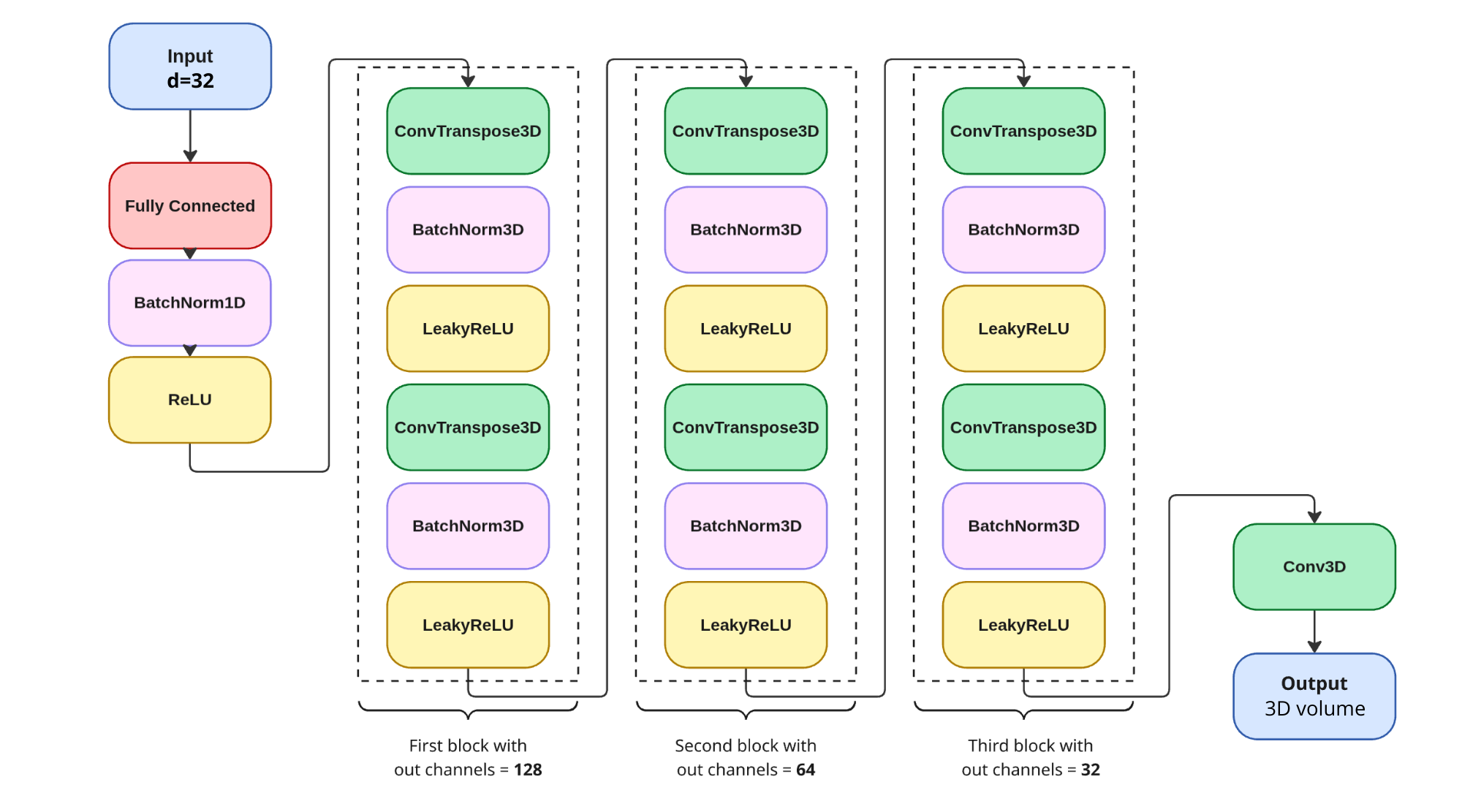}
\caption{Champollion's decoder backbone, largely inspired from Guillon et al.~\cite{Guillon2024}. It consists of six convolutional layers and is trained for 10 epochs with a Binary Cross Entropy loss, a learning rate of $5\times10^{-4}$, and a batch size of 32.}
\label{fig:decoder_backbone}
\end{figure}

\begin{figure}[p]
\centering
\includegraphics[width = 0.7\columnwidth]{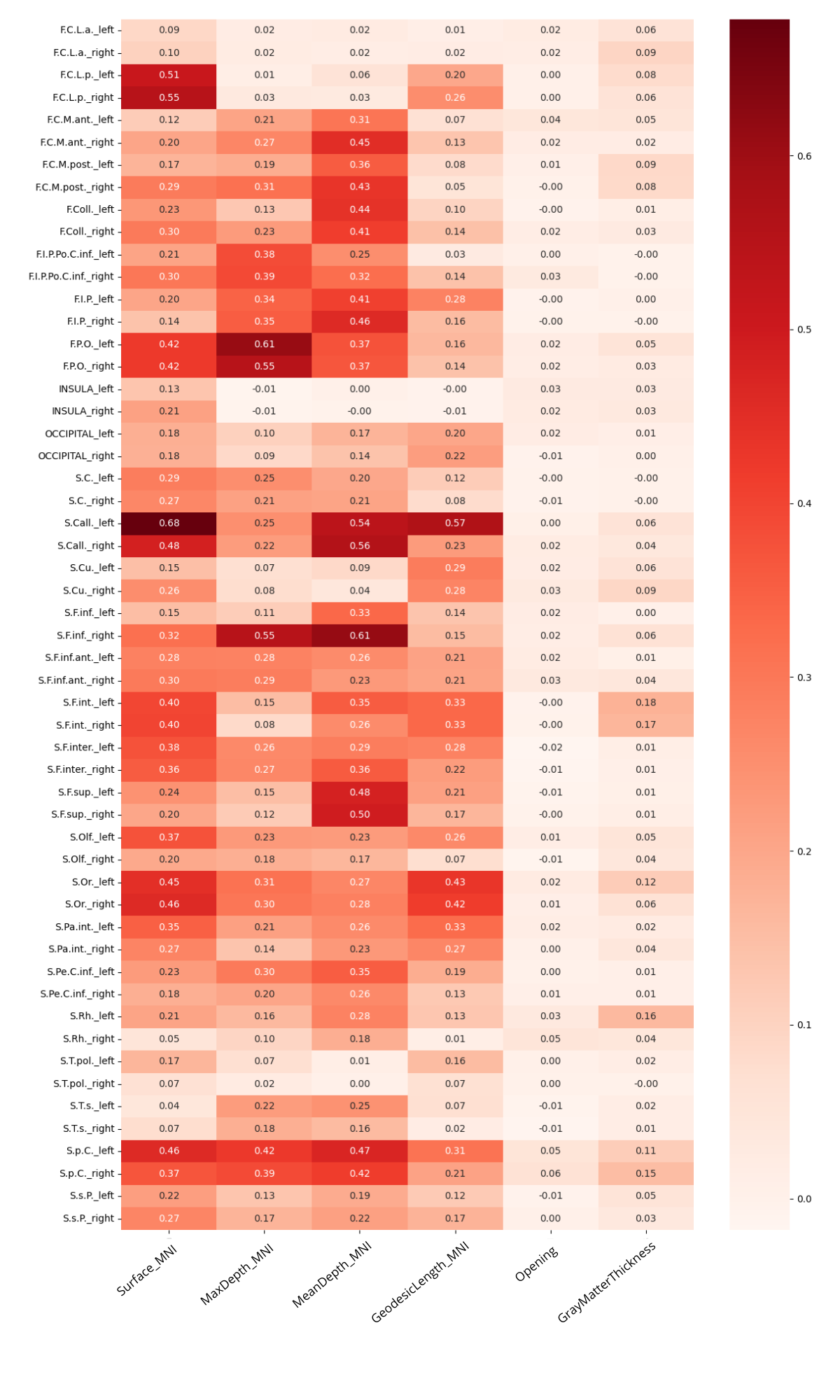}
\caption{Linear regression score (R²) of each morphometric descriptor of each sulcus automatically labeled by BrainVisa, based on Champollion's representation space. For each sulcus, the corresponding ROI to perform the regression is the smallest of the 56 ROIs containing the sulcus of interest. As expected, sulcal opening is not encoded by Champollion (mean $R^{2}=0.01$ across sulci, ranging in $[-0.02,0.06]$), since the skeleton representation is intrinsically independent of sulcal width. Likewise, gray matter thickness shows little to no correspondence with the representations (mean $R^{2}=0.04$, range $[0.00,0.18]$), consistent with the absence of cortical thickness information in the skeletons and suggesting that folding patterns are largely independent of cortical thickness. Conversely, morphometric descriptors directly related to sulcal shape are partially reflected in Champollion’s representation (geodesic length: mean $R^{2}=0.15$, range $[-0.01,0.57]$, maximum depth: mean $R^{2}=0.19$, range $[-0.01,0.61]$, mean depth: mean $R^{2}=0.21$, range $[0.00,0.61]$, and surface area: mean $R^{2}=0.24$, range $[0.04,0.68]$).}
\label{fig:morpho}
\end{figure}

\begin{figure}[p]
    \centering
    \begin{subfigure}{1.\textwidth}
        \includegraphics[width=1.\textwidth,trim={0 0 0 1.5cm},clip]{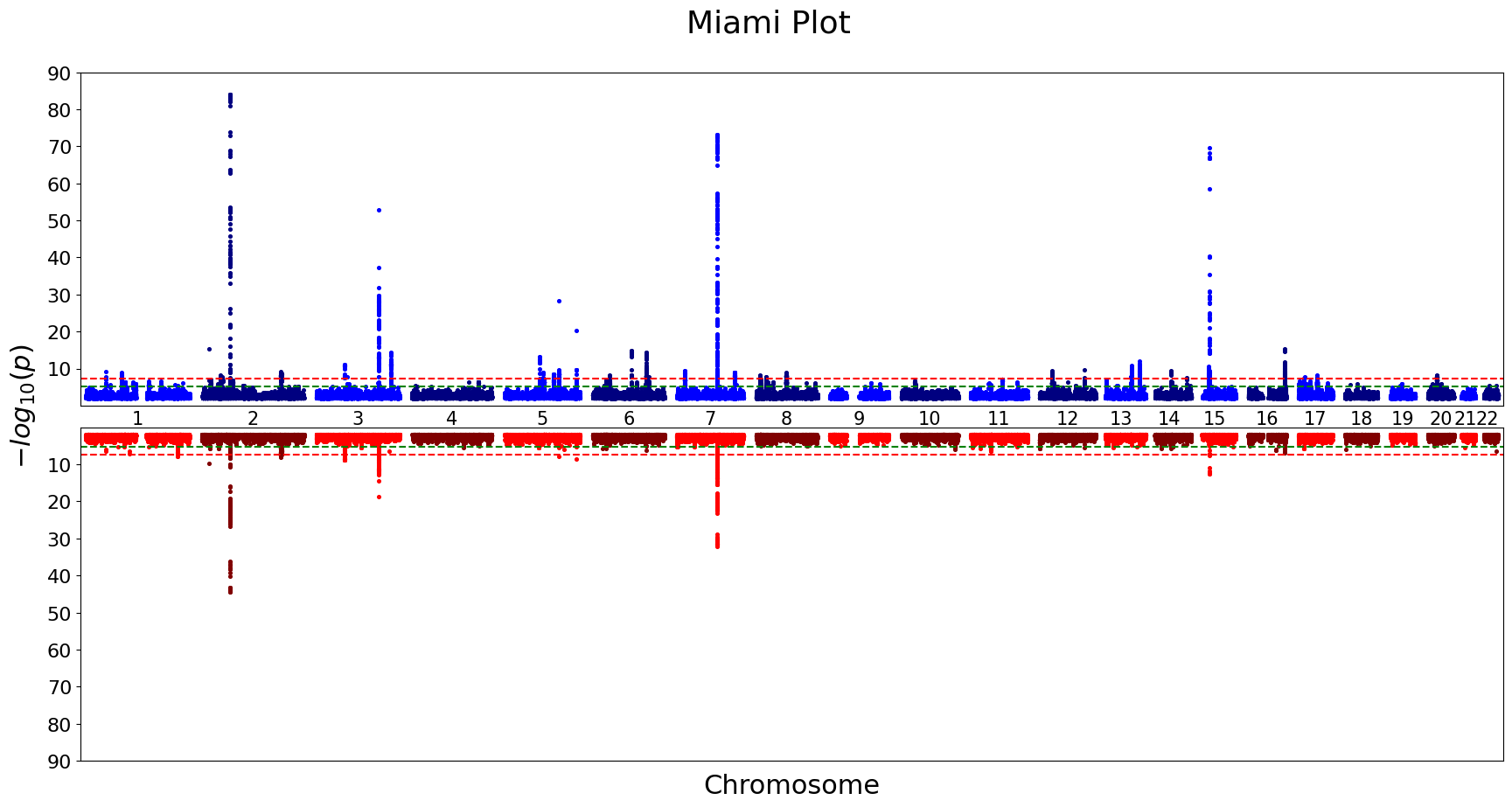}
        \caption{SFinf-BROCA-SPeCinf left region}
    \end{subfigure}
    \begin{subfigure}{1.\textwidth}
        \includegraphics[width=1.\textwidth,trim={0 0 0 1.5cm},clip]{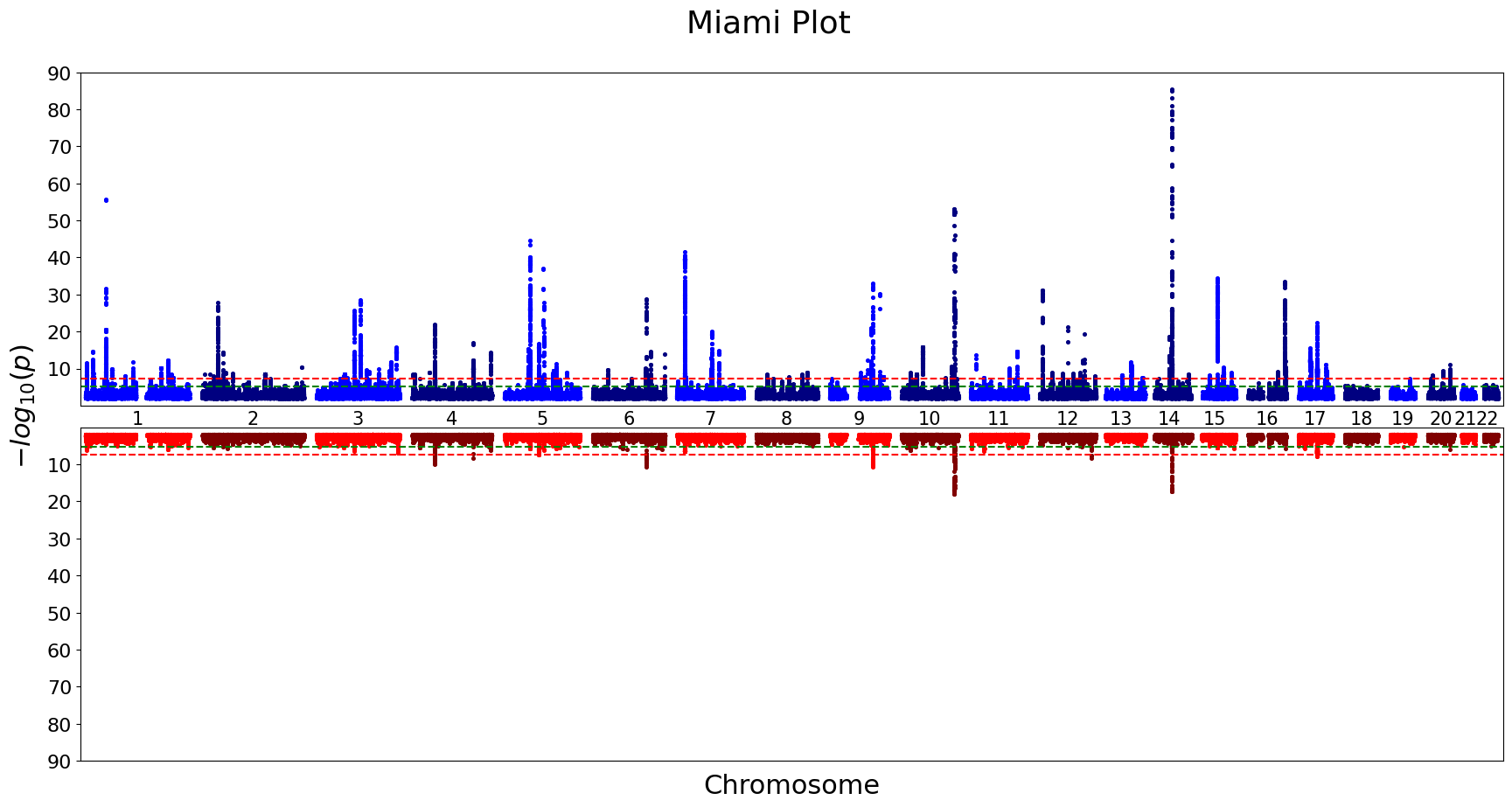}
        \caption{ScCal-SLi right region}
    \end{subfigure}
    \caption{Champollion representations encompass and exceed the genetic associations identified by classical sulcal morphometric descriptors. Miami plot comparing the associations using MOSTest (\textbf{a)} SFinf-BROCA-SPeCinf left and \textbf{b)} ScCal-SLi right encoded by Champollion on the $35{,}940$ UK BioBank white British ancestry subjects (blue) to the associations from the morphometric measurements (length of the sulci, surface of the sulci, mean depth, and max depth) (red). The conventional threshold of $5 \times 10^{-8}$ is indicated with the red line. The window is clipped for the $-\log_{10}(\text{p-value})$ above 90.}
    \label{fig:miami_plot}
\end{figure}

\begin{figure}[p]
\centering
\includegraphics[width = 0.8\columnwidth]{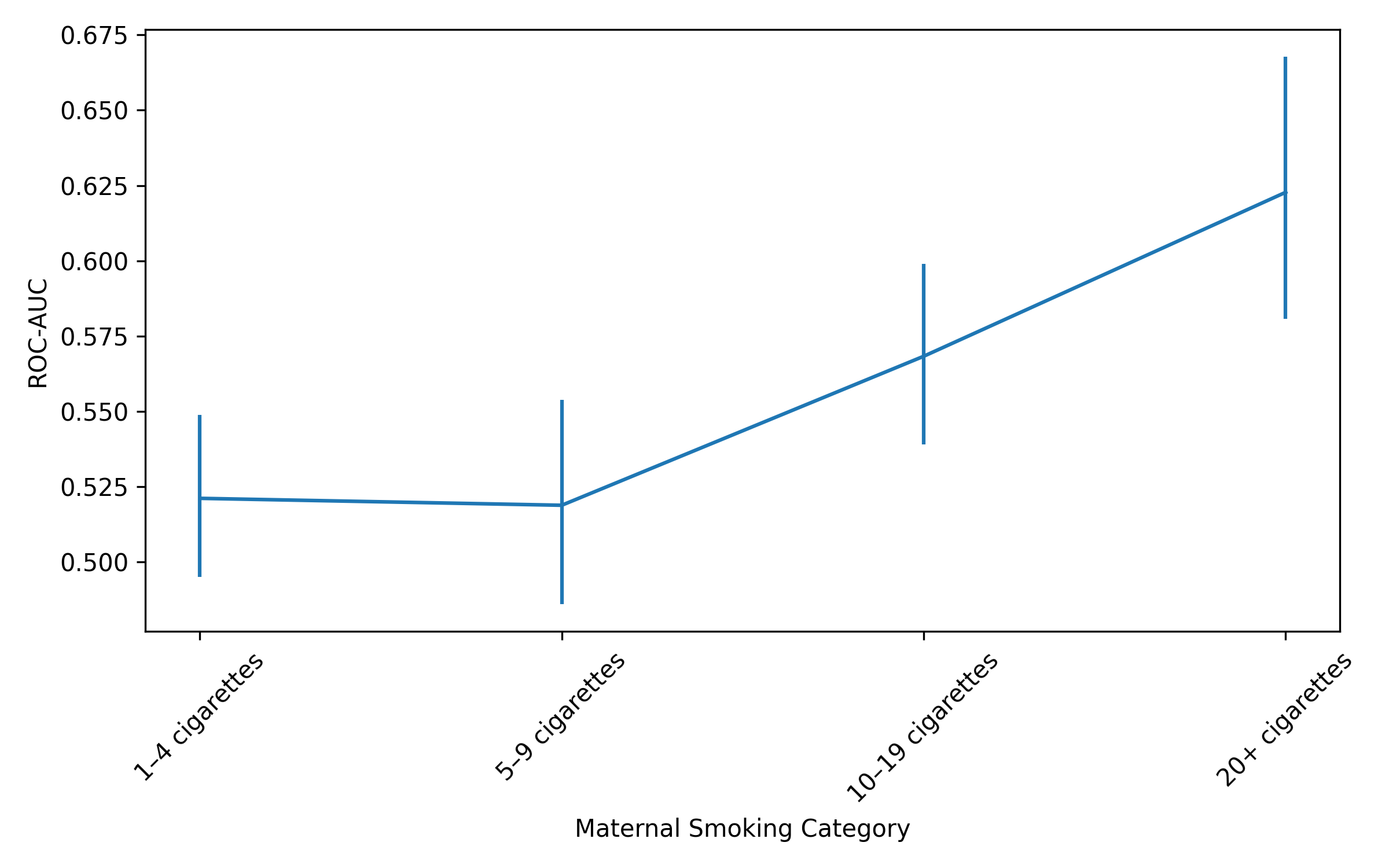}
\caption{Binary classification performance of maternal smoking exposure from the left occipital lobe representation space in the ABCD dataset, between controls and exposed subjects, for different exposure levels. The classification direction is fixed, obtained from a linear fit on UK BioBank data (yielding AUC=0.59 on UK BioBank). While a single binary variable is available on UK BioBank, the ABCD dataset provides a quantitative measurement of smoking exposure. Here, we not only show that the found signature is replicable across two different datasets presenting very different demographics (notably age), but the signature yields higher signal for higher smoking intensity (one-sided Spearman test among exposed subjects, excluding controls, yields $p=10^{-4}$, $N_{controls}=8{,}486$, $N_{1-4\,cigarettes}=376$, $N_{5-9\,cigarettes}=304$, $N_{10-19\,cigarettes}=384$, $N_{20+\,cigarettes}=134$). The error bars indicate the $95\%$ confidence intervals for each binary classification, estimated using stratified bootstrap resampling.}
\label{fig:abcd_smoking_auc}
\end{figure}

After each convolutional layer, we use batch normalization, LeakyReLu activation function, and a dropout of $5\%$. We use a learning rate of $4*10^{-4}$, and train for 80 epochs. A schematic of the architecture is provided in Extended Data, Figure~\ref{fig:backbone}.

Each ROI is modeled with its own backbone, resulting in $\sim100$ million convolutional parameters across the 56 regions that together cover the cortex.

\subsection{Optimization}

\textbf{General approach.} We optimized Champollion using a standard linear probing framework, in which representations are assessed via linear evaluation on sulcus-related classification and regression tasks. Importantly, we focused on tasks that directly reflect cortical folding patterns, rather than global biological traits such as sex, age, or pathology. This choice was motivated by the need to avoid overfitting on residual information (\textit{e.g.}, local brain volume or shape) that may be encoded residually after preprocessing. Optimizing on phenotypes whose primary sources of variance are unknown could encourage the model to rely on confounding factors rather than on folding-specific features. By contrast, restricting optimization to folding patterns ensures that the learned representations reflect the intended anatomical signal.

Because folding patterns are local, optimizing hyperparameters separately for each ROI would require annotations across many ROIs. This would be both impractical, given that folding patterns are sparsely studied and often unreported in several regions, and counterproductive, since we aim to keep the flexibility to define new ROIs without re-optimizing the model. Moreover, region-specific optimization risks local overfitting to a subset of patterns, potentially limiting the model’s ability to capture novel or unexplored configurations. To address this, we adopted a unified optimization strategy in which the same set of hyperparameters is applied across all regions, selected based on overall performance on a representative set of benchmark tasks.

We identified four downstream tasks that span both topological (sulcus presence, interruption types) and non-topological (shape regression) variability. These tasks were chosen from distinct cortical regions of varying size and complexity, as assessed by the number of folds and the diversity of observed configurations. They were also drawn from external datasets with substantial domain shifts (scanner type, resolution, and population), thereby encouraging robustness to site effects. Each selected pattern has previously been associated with behavioral or clinical phenotypes, reinforcing its biological relevance. Details of the tasks are provided below.

Although each pattern occurs bilaterally, we restricted optimization to one hemisphere per task to reduce computational demands. In practice, the final chosen configuration performed among the best across all four tasks, supporting its robustness and generalizability.

\textbf{Downstream tasks.} We used four downstream tasks, each associated with a specific sulcus, to guide the model optimization (Figure~\ref{fig:patterns}).

\begin{enumerate}
\item \textbf{Anterior cingulate cortex.} Binary classification of paracingulate sulcus presence in the anterior cingulate cortex (ACC) region ($N=341$, $42\%$ paracingulate positive), from the ACCpatterns dataset~\cite{chakravarty_striatal_2014,rapoport_childhood_2011,cachia_longitudinal_2016,delalande_complex_2020,tissier_sulcal_2018}. The paracingulate sulcus, when present, runs parallel to the cingulate sulcus, and its definition (including minimal length and allowable interruptions) follows~\cite{cachia_shape_2014}. Paracingulate sulcus asymmetry has previously been linked to control efficiency in preschoolers~\cite{cachia_shape_2014}.
\item \textbf{Orbito-frontal cortex.} Four-class classification of interruption types in the orbitofrontal cortex ($N=577$, Type I $49\%$, Type II $28\%$, Type III $17\%$, Type IV $6\%$), from the HCP dataset~\cite{van_essen_wu-minn_2013}, annotated by Troiani et al.~\cite{troiani_variability_2022}. The four interruption types correspond to distinct combinations of continuity and discontinuity across the medial orbital sulcus and lateral orbital sulcus. Orbitofrontal cortex interruption types have been linked to altered prevalence in schizophrenia and catatonia~\cite{isomura_altered_2017,moyal_orbitofrontal_2024}.
\item \textbf{Intra-parietal sulcus.} Binary classification of the intraparietal sulcus interruption ($N=390$, $69\%$ interrupted), from the HCP dataset, annotated by Santacroce et al.~\cite{Santacroce2024-dv}. The intraparietal sulcus morphology has been associated with memory and language performance.
\item \textbf{Central sulcus.} Regression of six continuous shape descriptors derived from the central sulcus in HCP ($N=880$). The descriptors were obtained using Isomap~\cite{tenenbaum_global_2000} on sulcal shape similarity matrices, following Sun et al.~\cite{sun_effect_2012}. The first dimension (the single- to double-knob pattern) has been linked to handedness~\cite{sun_effect_2012} and to functional variability in hand movement and reading~\cite{sun_linking_2016}. Of note, these descriptors do not constitute ground truth as they were obtained based on a computational method, but we considered reasonable to seek capturing as much Isomap information as possible. We restricted this approach to the central sulcus, given its reliable automatic identification, low interruption rate ($<1\%$), and clear anatomical boundaries. For other sulci, where automated labeling is less robust, label-agnostic methods such as Champollion may be better suited to capture continuous shape patterns than Isomap.
\end{enumerate}

\textbf{Choice of hemisphere.} For each task, we evaluated a single hemisphere during optimization to reduce computational requirements. Notably, for the orbitofrontal cortex, we used the left hemisphere, which displayed more balanced class distributions. For the intraparietal sulcus, we focused on the right hemisphere, as preliminary analysis suggested that it yielded more challenging predictions.

\textbf{Data splits.} For both ACCpatterns and HCP datasets, we used stratified data splits accounting for the label of interest, sex, age, and acquisition site. In the HCP dataset, siblings were assigned to the same split. Stratification was performed using \textit{StratifiedGroupKFold} from scikit-learn~\cite{scikit-learn}. For each task, we first partitioned 20\% of the subjects as a held-out test set, and then performed 5-fold stratified cross-validation on the remaining 80\% to optimize SSL hyperparameters. Grid search was conducted via \textit{GridSearchCV}, and the retained test sets were used exclusively for the final evaluation and baseline comparisons.

\textbf{Linear models.} Linear probing was implemented with \textit{LogisticRegression} for classification and \textit{ElasticNet} for regression (scikit-learn). For classification, hyperparameters included $C \in {10^{-3},10^{-2},10^{-1},1,10^{1},10^{2},10^{3}}$ and $\ell_1$ ratio $\in {0.1,\dots,1}$, using the saga solver with elastic-net penalty and a maximum of 2{,}000 iterations. For regression, we tuned $\ell_1$ ratio $\in {0.1,\dots,1}$ and $\alpha \in {10^{-3},10^{-2},10^{-1},1,10^{1},10^{2},10^{3}}$, with a maximum of 10{,}000 iterations.

\subsection{Evaluation of the baselines}

Champollion was benchmarked against multiple baselines on the same folding pattern downstream tasks. For each baseline, the same splits and linear models as for Champollion were used for hyperparameter search and final evaluation. \textit{Ridge} regression and \textit{RidgeClassifier} were used for the evaluation of the foundation models for very large representation space configurations, to reduce computation time.

First, a $\beta-VAE$ approach designed for anomaly detection on cortical folding skeletons~\cite{Guillon2024} was assessed. The model was originally trained on the HCP dataset (N=1114), much smaller than UK Biobank. We retrained the model on the same UK Biobank data as Champollion, and tuned it's representation space size in the range $[16,256]$ because it was designed for cortical areas of small complexity (\textit{e.g.} central sulcus and cingulate sulcus).

We then assessed a range of foundation models spanning a 2D natural imaging model (DINOv3~\cite{simeoni2025dinov3}), a point cloud model (Point-M2AE~\cite{zhang2022pointm2ae}, owing to its training on a large 3D shape dataset and the sparsity and binary nature of the cortical skeletons), and 3D medical models, designed for either representation (3DINO-ViT~\cite{xu2025threedino} and BrainSegFounder~\cite{cox2024brainsegfounder}, as we considered only its pretrained encoder), or segmentation (SAM-Med3D~\cite{wang2025sammed3d} and VISTA3D~\cite{he2025vista3d}). We included segmentation models because we believed they could offer spatially grounded representations and a good understanding of shapes. SAM-Med3D, VISTA3D and 3DINO-ViT were trained on multiple organs, including the brain. 3DINO-ViT and SAM-Med3D were trained on both MRI and CT scans, while VISTA3D was trained only on CT scans. BrainSegFounder was trained only on T1 and T2 MRI from UK Biobank.

To be as fair as possible with the models' capabilities to capture folding patterns in the zero-shot setting, we tested both ROI-centered MRI crops and cortical skeletons as input, with multiple preprocessing configurations to reduce the domain gap for skeletons, as well as various ways to extract representations. Overall, the benchmark study includes more than $1{,}500$ tested configurations and is therefore detailed in Supplementary (section 4).

The best performing foundation model in the zero-shot setting, namely 3DINO-ViT, was then adapted to cortical skeletons using continual pretraining with LoRA~\cite{hu2022lora}, incorporating a masking strategy adapted to the sparsity and binary nature of the cortical skeletons, as well as the positional augmentations used in Champollion (Supplementary sections 4.1.6 and 4.2.1).


\subsection{Multivariate genetics associations}

For a given ROI, two types of multivariate phenotypes were compared. The first is the 32-dimensional representation space from Champollion. The second is a conventional tabular representation of the ROI built from sulcal morphometric measures (geodesic length, surface area, mean depth, and maximum depth).

A given ROI always encompasses the sulci indicated by its name. For instance, the region SC-SPeC left encompasses the central sulcus (SC) and the precentral sulcus (SPeC). Following the BrainVISA nomenclature, an automated deep learning-based sulcal labeling model from Morphologist~v5.0 was used to segment cortical folding skeletons into individual sulci. A first threshold was applied to ensure sufficient sample size: only sulci present in more than 90\% of the individuals were retained. For each labeled sulcus, the following morphometric parameters were then extracted, after affine alignment in the MNI spatial referential: geodesic length at the junction with the cortical hull, surface area, mean depth, and maximum depth.

To ensure consistency with the Champollion regional framework, sulcal morphometric parameters were concatenated according to the same regional definitions. Thus, the morphometric characterization of a given region is based on the set of length, surface, and depth measures associated with all the sulci composing that region.

Pre-residualization was performed on both the representations and the tabular morphometric data. For the UK Biobank cohort, the following covariates were included: sex (field~31), age at imaging (field~53), age$^{2}$, age $\times$ sex and age$^{2}$ $\times$ sex interactions, imaging center (field~54), total intracranial volume (TIV) using \texttt{CAT12} (v12.8), and the first 20 principal genetic components (field~22009).

Following Dufournet et al.~\cite{Dufournet2025-sh}, multivariate genetic associations were assessed using the Multivariate Omnibus Statistical Test (MOSTest), which combines summary statistics from univariate genome-wide association studies (GWAS) conducted on each phenotype~\cite{van2020understanding}, accounting for the correlation structure between univariate phenotypes. MOSTest was applied separately to each of the two multivariate phenotypes, the representation space and the tabular morphometric data, yielding two sets of genome-wide association results per ROI. Independent genomic loci were then defined for each result using the Functional Mapping and Annotation of Genome-Wide Association Studies (FUMA) platform~\cite{watanabe2017functional}. For a given ROI, the overlap between the two sets of loci was quantified as the number of loci identified from the tabular morphometric data whose genomic coordinates (start and end positions) intersect with at least one locus identified from the representation space.

To reduce the dimensional imbalance between classical morphometry and Champollion’s 32-dimensional representation space, we selected ROIs maximizing both the number of retained subjects and the number of sulci, thereby increasing the number of classical morphometric phenotypes. Specifically, we selected one ROI with weak-to-moderate correlations between morphometry and Champollion's representation (ScCal–SLi\_right, 12 classical morphometric phenotypes in 31,606 subjects) and one with strong correlations (SFinf–BROCA–SPeCinf\_left, 20 phenotypes in 29,944 subjects). Furthermore, the multivariate genetic framework explicitly accounts for the number of phenotypes when assessing statistical significance.

\subsection{Exploratory analysis}

All classifications were performed by linear probing of Champollion’s representations, after site harmonization with ComBat~\cite{fortin2018harmonization,johnson2007adjusting} (using \texttt{NeuroCombat-sklearn}) and residualization on sex, age, and socio-economic status (when available).

More specifically, for the left Incomplete Hippocampal Inversion (IHI) classification, we used the QTIM dataset~\cite{qtim,strike_genetic_2019}: there is a single recording site. Variables representing socio-economic status are not available. During fold stratification, we ensured that the twins were in the same fold to avoid data leakage. We took as a measure of the IHI the global assessment of the presence of a left IHI, C0, determined by Cury et al.\cite{cury2015incomplete}, considering only the two extreme levels for classification: pronounced IHI or absence of the IHI ($N=928$, $8\%$ pronounced IHI).

For prematurity classification, we used the ABCD dataset~\cite{volkow2018conception}. We considered only the very preterms (born after 28-32 weeks post-menstrual age, according to the World Health Organization definition~\cite{WHO_preterm}, also called great prematurity), classified against the full terms ($>37$ weeks post-menstrual age, $N=8{,}846$, $1\%$ preterms).

For maternal smoking exposure during pregnancy, we used the UK Biobank dataset~\cite{sudlow2015uk} and the ABCD dataset for replication. UK BioBank was used to fit the classifier, based on binary label reported by the offspring. No site harmonization was performed since the 3 sites are homogeneous (same scanner, same protocol). ABCD contains quantitative labels (categories of average number of cigarettes per day), and the UK BioBank classifier was applied to the different categories against controls.

The predictive tasks consisted of binary classifications across ROI representations, and each model performance was quantified using ROC-AUC. Classification models were implemented using \texttt{scikit-learn} logistic regression with an L2 penalty, the \textit{lbfgs} solver, balanced class weights, and a maximum of $1{,}000$ iterations. The regularization parameter $C$ was selected via 5-fold cross-validation from a predefined grid $C \in \{10^k: k = -7, \ldots, 3\}$. Within each cross-validation fold, predictors were residualized for confounding variables using regression models fitted on the training data only, and the resulting residuals were applied to both the training and test sets. Residualization was implemented using \texttt{statsmodels}, allowing appropriate handling of both categorical and continuous confounders through the construction of design matrices.

Statistical significance was assessed using a permutation test based on the \emph{ter Braak scheme}~\cite{ter_braak_permutation_1992}, which tests for associations between predictors and outcome conditional on confounders. Permutation tests used a unique regularization coefficient, selected on the basis of the non-permuted cross-validation. Under the null hypothesis, only the residual component of the predictors is exchangeable; accordingly, permutation was applied solely to the residualized predictors in the training set. This procedure preserves the relationships between predictors and confounders, as well as between confounders and the outcome, while breaking any association between predictors and outcome beyond confounders. Cross-validation was embedded within the permutation procedure to avoid information leakage and to obtain a valid empirical null distribution for the AUC. To account for the number of statistical tests, we set the significance threshold to $\frac{0.05}{56}$ (Bonferroni correction). Accordingly, the number of permutation tests per ROI was set to $10\times\frac{56}{0.05} = 11{,}200$.

For each task, the best performing ROI was then selected for pattern interpretation through latent traversal along the classification direction, orthogonally to the associated hyperplane. Because classification directions typically explain only a small fraction of the total variance, latent traversals induce subtle but structured changes. Reliable interpretation therefore relies on identifying consistent variations across multiple subjects. Accordingly, the top 10 best predicted positive and negative subjects were sampled for latent traversal and associated visual interpretation (5 of which are shown in Supplementary Figures 15-19). The subjects were decoded at their original position in the representation space, and after translation to reach the median of the univariate distribution, and the opposit end (quantile 0.99 for originally negative samples, and quantile 0.01 for positive samples; distribution quantiles were calculated based on UK BioBank.)

\begin{figure}[p]
\centering
\includegraphics[width = 1.\columnwidth]{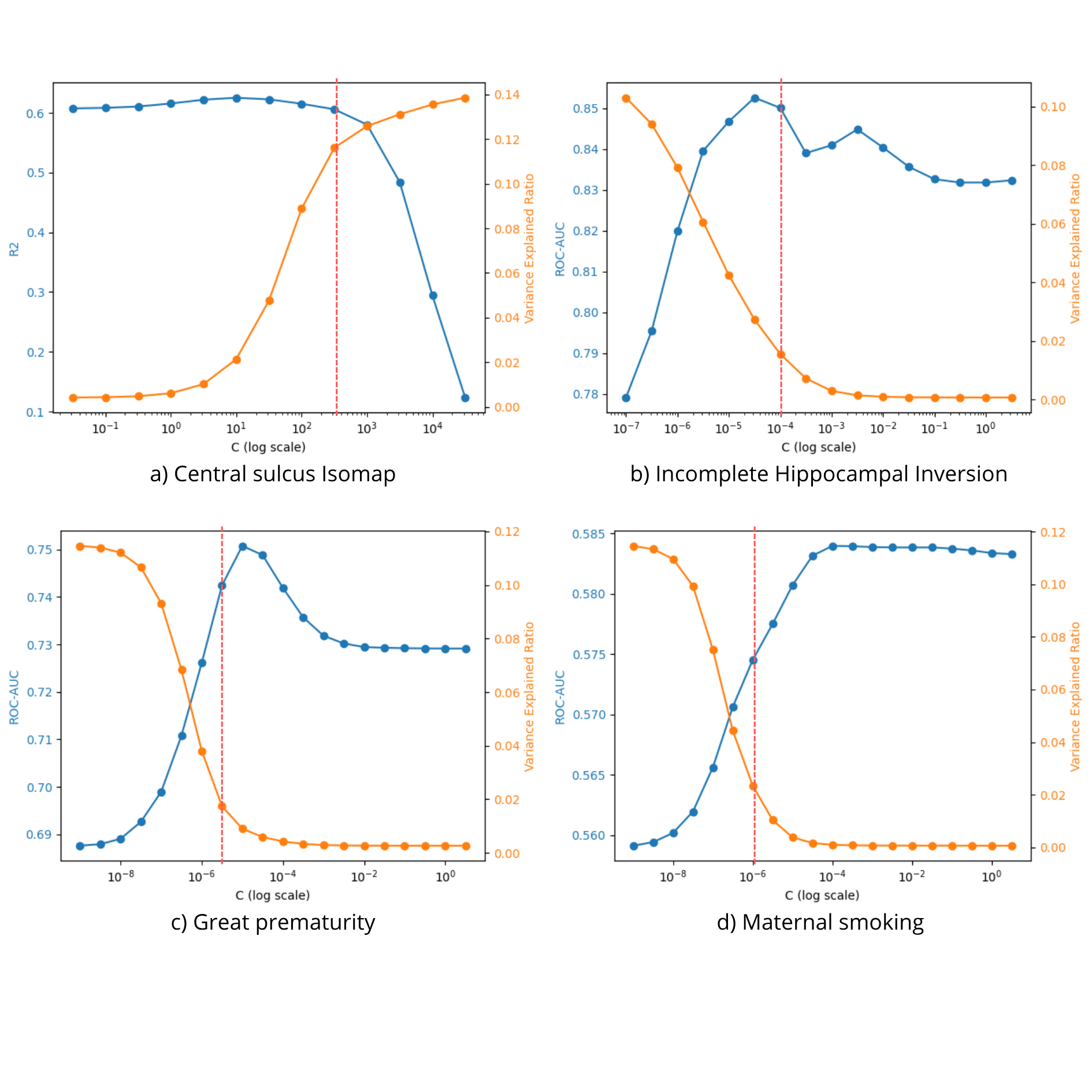}
\caption{For each exploratory analysis task, the latent direction obtained via linear probing of the highest scoring ROI is used for latent traversal. Selecting the regularization parameter of the linear classifier / regressor on a performance basis only can lead to directions of small explained variance, limiting the visual interpretability of the latent traversal. By drawing the performance against the explained variance for different directions obtained with different regularization coefficients, we select a direction that maximizes performance under an explained variance threshold constraint ($\sim 2\%$), which is empirically found to yield sufficient visual signal, while barely affecting performance ($<2\%$). The selected coefficient is represented by the red dotted line.}
\label{explained_variance}
\end{figure}

Furthermore, to ensure sufficient interpretability, we enforced the classification direction to explain at least $2\%$ of the global variance of the dataset (computed as for PCA dimensions, $\frac{v^T\Sigma v}{Tr(\Sigma)}$, where $v$ is the direction and $\Sigma$ the covariance matrix). Indeed, we empirically observed that directions of larger variance typically correspond to a larger number of affected voxels in the reconstructed space. We found that $2\%$ of the global variance always yielded a signal strong enough to be visually interpreted through the latent traversal, while barely affecting the classifier's performance ($<2\%$ on all investigated tasks). In practice, by applying different regularization strengths in the Logistic Regression on the non-standardized representation space, we can draw a distribution of the classifier's performance depending on the explained variance ratio of the corresponding direction. Stronger regularization implicitly enforces a higher variance ratio. For each task, the regularization coefficient and the corresponding classification direction were chosen based on such a curve, reported in Extended Data Figure~\ref{explained_variance}.

We additionally assessed the stability of the learned direction using bootstrap resampling, reporting the mean pairwise Pearson correlation of prediction scores across runs. To relate the obtained quantitative values to possible instabilities in visual interpretation, we selected the least stable task of the exploratory analysis (prematurity) and one representative direction obtained under bootstrap (sampled at median correlation with the original direction across bootstrap runs). The two directions show near perfect visual correspondence, confirming stability for the tasks under study (Supplementary Figures 17-18).

\section{Data availability}

This research has been conducted using the UK Biobank Resource~\cite{sudlow2015uk} under Application Number \textbf{64984}.

Data used for the prematurity study in the preparation of this article were obtained from the Adolescent Brain Cognitive Development™ (ABCD) Study~\url{https://abcdstudy.org/}, held in the NIH Brain Development Cohorts Data Sharing Platform~\url{https://www.nbdc-datahub.org/}. This is a multisite, longitudinal study designed to recruit more than \num{10000} children aged 9-10 and follow them over 10 years into early adulthood. The ABCD Study is supported by the National Institutes of Health and additional federal partners under award numbers: U01DA041048, U01DA050989, U01DA051016, U01DA041022, U01DA051018, U01DA051037, U01DA050987, U01DA041174,\\
U01DA041106, U01DA041117, U01DA041028, U01DA041134, U01DA050988, U01DA051039,\\
U01DA041156, U01DA041025,U01DA041120, U01DA051038, U01DA041148, U01DA041093,\\
U01DA041089, U24DA041123, U24DA041147. A full list of supporters is available at Federal Partners - ABCD Study~\url{https://abcdstudy.org/federal-partners.html}. ABCD Consortium investigators designed and implemented the study and/or provided data but did not necessarily participate in the analysis or writing of this report. This manuscript reflects the views of the authors and may not reflect the opinions or views of the NIH or ABCD Consortium investigators.

MRI data for the QTIM study used for incomplete hippocampal inversion classification are publicly available for download~\cite{qtim}.

Data for the twins study used in the preparation of this article were obtained from the Human Connectome Project – Young Adult (HCP YA) database (humanconnectome.org)~\cite{van_essen_wu-minn_2013}. Data were provided [in part] by the Human Connectome Project, WU-Minn Consortium (Principal Investigators: David Van Essen and Kamil Ugurbil; 1U54MH091657) funded by the 16 NIH Institutes and Centers that support the NIH Blueprint for Neuroscience Research; and by the McDonnell Center for Systems Neuroscience at Washington University. 

\section{Code availability}
The preprocessing toolbox is available on Github (Cortical tiles: \url{https://github.com/neurospin/cortical_tiles}). The model weights and embedding generation pipeline are available on HuggingFace (Champollion: \url{https://huggingface.co/neurospin/Champollion_V1}). A full processing from MRI to embeddings demo is also available on HuggingFace, where users can upload a few MRIs to see whether their data is properly processed, without any installation required (Demo: \url{https://huggingface.co/spaces/neurospin/Champollion_demo}). The complete pipeline, from MRI to Morphologist cortical graphs to Champollion embeddings, is available at \url{https://github.com/neurospin/champollion_pipeline}.
Additional codes are available in the following repositories:
\begin{itemize}[noitemsep]
\item Champollion decoder: \url{https://github.com/neurospin-projects/2025_adufournet_champollion_decoder}
\item Post-GWAS tools for genetic analysis: \url{https://github.com/neurospin/postgwas-tools}
\item Evaluation of foundation models: \url{https://github.com/neurospin/champollion_vs_foundation_models}
\end{itemize}

\printbibliography

@ARTICLE{Cachia2021-vb,
  title     = "Towards deciphering the fetal foundation of normal cognition and
               cognitive symptoms from sulcation of the cortex",
  author    = "Cachia, Arnaud and Borst, Gr{\'e}goire and Jardri, Renaud and
               Raznahan, Armin and Murray, Graham K and Mangin, Jean-Fran{\c
               c}ois and Plaze, Marion",
  journal   = "Front. Neuroanat.",
  publisher = "Frontiers Media SA",
  volume    =  15,
  pages     = "712862",
  month     =  sep,
  year      =  2021,
  copyright = "https://creativecommons.org/licenses/by/4.0/",
  language  = "en"
}

@article{Snyder2026,
  title = {Genetic insights on the mechanisms of human cortical folding},
  url = {http://dx.doi.org/10.64898/2026.03.06.709690},
  DOI = {10.64898/2026.03.06.709690},
  journal = {bioRxiv},
  publisher = {openRxiv},
  author = {Snyder,  William and Shafee,  Rebecca and Liu,  Siyuan and Levitis,  Elizabeth and Duan,  Kuaikuai and Kumar,  Kuldeep and Schleifer,  Charles H and Boen,  Rune and Ching,  Christopher RK and Han,  Joan C. and Lee,  Nancy and Mulle,  Jennifer G and Shultz,  Sarah and Jacquemont,  Sébastien and Bearden,  Carrie E and Vértes,  Petra E and Bullmore,  Edward T and Raznahan,  Armin},
  year = {2026},
  month = Mar 
}

@ARTICLE{Regis2011-kq,
  title     = "Subclinical abnormal gyration pattern, a potential anatomic
               marker of epileptogenic zone in patients with magnetic resonance
               imaging-negative frontal lobe epilepsy",
  author    = "R{\'e}gis, Jean and Tamura, Manabu and Park, Michael C and
               McGonigal, Aileen and Rivi{\`e}re, Denis and Coulon, Olivier and
               Bartolomei, Fabrice and Girard, Nadine and Figarella-Branger,
               Dominique and Chauvel, Patrick and Mangin, Jean-Fran{\c c}ois",
  journal   = "Neurosurgery",
  publisher = "Ovid Technologies (Wolters Kluwer Health)",
  volume    =  69,
  number    =  1,
  pages     = "80--93; discussion 93--4",
  month     =  jul,
  year      =  2011,
  language  = "en"
}

@ARTICLE{Fischl2008-nj,
  title     = "Cortical folding patterns and predicting cytoarchitecture",
  author    = "Fischl, Bruce and Rajendran, Niranjini and Busa, Evelina and
               Augustinack, Jean and Hinds, Oliver and Yeo, B T Thomas and
               Mohlberg, Hartmut and Amunts, Katrin and Zilles, Karl",
  journal   = "Cereb. Cortex",
  publisher = "Oxford University Press (OUP)",
  volume    =  18,
  number    =  8,
  pages     = "1973--1980",
  month     =  aug,
  year      =  2008,
  language  = "en"
}

@inproceedings{hu2022lora,
title={Lo{RA}: Low-Rank Adaptation of Large Language Models},
author={Edward J Hu and Yelong Shen and Phillip Wallis and Zeyuan Allen-Zhu and Yuanzhi Li and Shean Wang and Lu Wang and Weizhu Chen},
booktitle={International Conference on Learning Representations},
year={2022},
url={https://openreview.net/forum?id=nZeVKeeFYf9}
}

@article{SimSiamCollapse,
     title={Understanding Collapse in Non-Contrastive 
            Siamese Representation Learning},
     author={Li, Alexander Cong and Efros, Alexei A. and Pathak, Deepak},
     journal={ECCV},     
     year={2022}
}

@article{liu2022bridging,
  title={Bridging the gap from asymmetry tricks to decorrelation principles in non-contrastive self-supervised learning},
  author={Liu, Kang-Jun and Suganuma, Masanori and Okatani, Takayuki},
  journal={Advances in Neural Information Processing Systems},
  volume={35},
  pages={19824--19835},
  year={2022}
}

@Article{chen2020simsiam,
  author  = {Xinlei Chen and Kaiming He},
  title   = {Exploring Simple Siamese Representation Learning},
  journal = {arXiv preprint arXiv:2011.10566},
  year    = {2020},
}

@ARTICLE{Borne2020-fp,
  title     = "Automatic labeling of cortical sulci using patch- or {CNN-based}
               segmentation techniques combined with bottom-up geometric
               constraints",
  author    = "Borne, L{\'e}onie and Rivi{\`e}re, Denis and Mancip, Martial and
               Mangin, Jean-Fran{\c c}ois",
  journal   = "Med. Image Anal.",
  publisher = "Elsevier BV",
  volume    =  62,
  number    =  101651,
  pages     = "101651",
  month     =  may,
  year      =  2020,
  copyright = "http://creativecommons.org/licenses/by-nc-nd/4.0/",
  language  = "en"
}

@ARTICLE{Mangin1995-qp,
  title     = "From {3D} magnetic resonance images to structural
               representations of the cortex topography using topology
               preserving deformations",
  author    = "Mangin, Jean-François and Frouin, Vincent and Bloch, Isabelle
               and Régis, Jean and Lopez-Krahe, Jaime",
  journal   = "J. Math. Imaging Vis.",
  publisher = "Springer Science and Business Media LLC",
  volume    =  5,
  number    =  4,
  pages     = "297--318",
  month     =  dec,
  year      =  1995,
  language  = "en"
}

@misc{WHO_preterm,
  author       = {{World Health Organization}},
  title        = {Preterm birth},
  year         = {2026},
  howpublished = {\url{https://www.who.int/news-room/fact-sheets/detail/preterm-birth}},
}

@ARTICLE{mangin_tmi,
  author={Mangin, J.-F. and Riviere, D. and Cachia, A. and Duchesnay, E. and Cointepas, Y. and Papadopoulos-Orfanos, D. and Collins, D.L. and Evans, A.C. and Regis, J.},
  journal={IEEE Transactions on Medical Imaging}, 
  title={Object-based morphometry of the cerebral cortex}, 
  year={2004},
  volume={23},
  number={8},
  pages={968-982},
  doi={10.1109/TMI.2004.831204}}

@article{simeoni2025dinov3,
  title={Dinov3},
  author={Sim{\'e}oni, Oriane and Vo, Huy V and Seitzer, Maximilian and Baldassarre, Federico and Oquab, Maxime and Jose, Cijo and Khalidov, Vasil and Szafraniec, Marc and Yi, Seungeun and Ramamonjisoa, Micha{\"e}l and others},
  journal={arXiv preprint arXiv:2508.10104},
  year={2025}
}

@InProceedings{wang2025sammed3d,
author="Wang, Haoyu
and Guo, Sizheng
and Ye, Jin
and Deng, Zhongying
and Cheng, Junlong
and Li, Tianbin
and Chen, Jianpin
and Su, Yanzhou
and Huang, Ziyan
and Shen, Yiqing
and Fu, Bin
and Zhang, Shaoting
and He, Junjun
and Qiao, Yu",
editor="Del Bue, Alessio
and Canton, Cristian
and Pont-Tuset, Jordi
and Tommasi, Tatiana",
title="SAM-Med3D: Towards General-Purpose Segmentation Models for Volumetric Medical Images",
booktitle="Computer Vision -- ECCV 2024 Workshops",
year="2025",
publisher="Springer Nature Switzerland",
address="Cham",
pages="51--67",
isbn="978-3-031-91721-9"
}

@article{cox2024brainsegfounder,
  title   = {{BrainSegFounder}: Towards 3D Foundation Models for Neuroimage Segmentation},
  author  = {Cox, Joseph and Liu, Peng and Stolte, Skylar E. and Yang, Yunchao and Liu, Kang and See, Kyle B. and Ju, Huiwen and Fang, Ruogu},
  journal = {Medical Image Analysis},
  volume  = {97},
  pages   = {103301},
  year    = {2024},
  doi     = {10.1016/j.media.2024.103301}
}

@inproceedings{he2025vista3d,
  title     = {{VISTA3D}: A Unified Segmentation Foundation Model for 3D Medical Imaging},
  author    = {He, Yufan and Guo, Pengfei and Tang, Yucheng and Myronenko, Andriy and Nath, Vishwesh and Xu, Ziyue and Yang, Dong and Zhao, Can and Simon, Benjamin and Belue, Mason and Harmon, Stephanie and Turkbey, Baris and Xu, Daguang and Li, Wenqi},
  booktitle = {Proceedings of the IEEE/CVF Conference on Computer Vision and Pattern Recognition (CVPR)},
  pages     = {20863--20873},
  year      = {2025}
}

@article{xu2025threedino,
  title   = {A Generalizable 3D Framework and Model for Self-Supervised Learning in Medical Imaging},
  author  = {Xu, Tony and Hosseini, Sepehr and Anderson, Chris and Rinaldi, Anthony and Krishnan, Rahul G. and Martel, Anne L. and Goubran, Maged},
  journal = {npj Digital Medicine},
  volume  = {8},
  pages   = {639},
  year    = {2025},
  doi     = {10.1038/s41746-025-02035-w}
}

@inproceedings{zhang2022pointm2ae,
 author = {Zhang, Renrui and Guo, Ziyu and Gao, Peng and Fang, Rongyao and Zhao, Bin and Wang, Dong and Qiao, Yu and Li, Hongsheng},
 booktitle = {Advances in Neural Information Processing Systems},
 editor = {S. Koyejo and S. Mohamed and A. Agarwal and D. Belgrave and K. Cho and A. Oh},
 pages = {27061--27074},
 publisher = {Curran Associates, Inc.},
 title = {Point-M2AE: Multi-scale Masked Autoencoders for Hierarchical Point Cloud Pre-training},
 url = {https://proceedings.neurips.cc/paper_files/paper/2022/file/ad1d7a4df30a9c0c46b387815a774a84-Paper-Conference.pdf},
 volume = {35},
 year = {2022}
}

@ARTICLE{Pizzagalli2020-qy,
  title     = "The reliability and heritability of cortical folds and their
               genetic correlations across hemispheres",
  author    = "Pizzagalli, Fabrizio and Auzias, Guillaume and Yang, Qifan and
               Mathias, Samuel R and Faskowitz, Joshua and Boyd, Joshua D and
               Amini, Armand and Rivi{\`e}re, Denis and McMahon, Katie L and de
               Zubicaray, Greig I and Martin, Nicholas G and Mangin,
               Jean-Fran{\c c}ois and Glahn, David C and Blangero, John and
               Wright, Margaret J and Thompson, Paul M and Kochunov, Peter and
               Jahanshad, Neda",
  journal   = "Commun. Biol.",
  publisher = "Springer Science and Business Media LLC",
  volume    =  3,
  number    =  1,
  pages     = "510",
  month     =  sep,
  year      =  2020,
  copyright = "https://creativecommons.org/licenses/by/4.0",
  language  = "en"
}

@article{chi1977gyral,
  title={Gyral development of the human brain},
  author={Chi, Je G and Dooling, Elizabeth C and Gilles, Floyd H},
  journal={Annals of Neurology: Official Journal of the American Neurological Association and the Child Neurology Society},
  volume={1},
  number={1},
  pages={86--93},
  year={1977},
  publisher={Wiley Online Library}
}

@article{barbano2025anatomical,
  title={Anatomical foundation models for brain MRIs},
  author={Barbano, Carlo Alberto and Brunello, Matteo and Dufumier, Benoit and Grangetto, Marco and Alzheimer’s Disease Neuroimaging Initiative and others},
  journal={Pattern Recognition Letters},
  year={2025},
  publisher={Elsevier}
}

@ARTICLE{Perrot2011-pm,
  title     = "Cortical sulci recognition and spatial normalization",
  author    = "Perrot, Matthieu and Rivi{\`e}re, Denis and Mangin, Jean-Fran{\c
               c}ois",
  journal   = "Med. Image Anal.",
  publisher = "Elsevier BV",
  volume    =  15,
  number    =  4,
  pages     = "529--550",
  month     =  aug,
  year      =  2011,
  language  = "en"
}

@ARTICLE{Evans1992-bd,
  title     = "Anatomical mapping of functional activation in stereotactic
               coordinate space",
  author    = "Evans, A C and Marrett, S and Neelin, P and Collins, L and
               Worsley, K and Dai, W and Milot, S and Meyer, E and Bub, D",
  journal   = "Neuroimage",
  publisher = "Elsevier BV",
  volume    =  1,
  number    =  1,
  pages     = "43--53",
  month     =  aug,
  year      =  1992,
  language  = "en"
}

@ARTICLE{Dufournet2025-sh,
  title     = "A self-supervised learning framework for discovering cortical
               folding patterns under genetic influence: Application to the
               Anterior Cingulate Cortex",
  author    = "Dufournet, Antoine and Laval, Julien and Rivi{\`e}re, Denis and
               de Vareilles, H{\'e}lo{\"\i}se and Murray, Graham K and Cachia,
               Arnaud and Chavas, Jo{\"e}l and Frouin, Vincent and Mangin,
               Jean-Fran{\c c}ois",
  journal   = "Imaging Neurosci. (Camb.)",
  publisher = "MIT Press",
  volume    =  3,
  number    = "IMAG.a.987",
  pages     = "IMAG.a.987",
  month     =  nov,
  year      =  2025,
  copyright = "https://creativecommons.org/licenses/by/4.0/",
  language  = "en"
}

@article{ashburner2000voxel,
  title={Voxel-based morphometry—the methods},
  author={Ashburner, John and Friston, Karl J},
  journal={Neuroimage},
  volume={11},
  number={6},
  pages={805--821},
  year={2000},
  publisher={Elsevier}
}

@ARTICLE{Mancuso2025-gl,
  title     = "Functional connectivity of the superior temporal sulcus at
               term-equivalent age: effects of gestational age and sex",
  author    = "Mancuso, Charlotte and Bacquet, Maxime and Benjamin, Lucas and
               Leroy, Fran{\c c}ois and Dehaene-Lambertz, Ghislaine",
  journal   = "Brain Struct. Funct.",
  publisher = "Springer Science and Business Media LLC",
  volume    =  230,
  number    =  7,
  pages     = "123",
  month     =  jul,
  year      =  2025,
  copyright = "https://creativecommons.org/licenses/by/4.0",
  language  = "en"
}

@ARTICLE{Benjamin2025-zp,
  title     = "The auditory environment drives superior temporal sulcus depth
               in the neonatal period",
  author    = "Benjamin, Lucas and Bacquet, Maxime and Leroy, Fran{\c c}ois and
               Mancuso, Charlotte and Lordier, Lara and Sa de Almeida, Joana
               and Gui, Laura and Lean, Rachel E and Rogers, Cynthia E and
               Inder, Terrie and Smyser, Christopher D and H{\"u}ppi, Petra S
               and Dehaene-Lambertz, Ghislaine",
  journal   = "Brain Struct. Funct.",
  publisher = "Springer Science and Business Media LLC",
  volume    =  230,
  number    =  9,
  pages     = "177",
  month     =  nov,
  year      =  2025,
  copyright = "https://creativecommons.org/licenses/by-nc-nd/4.0",
  language  = "en"
}

@ARTICLE{Fischl2012-fy,
  title     = "{FreeSurfer}",
  author    = "Fischl, Bruce",
  journal   = "Neuroimage",
  publisher = "Elsevier BV",
  volume    =  62,
  number    =  2,
  pages     = "774--781",
  month     =  aug,
  year      =  2012,
  language  = "en"
}

@ARTICLE{El_Marroun2014-bv,
  title     = "Prenatal tobacco exposure and brain morphology: a prospective
               study in young children",
  author    = "El Marroun, Hanan and Schmidt, Marcus N and Franken, Ingmar H A
               and Jaddoe, Vincent W V and Hofman, Albert and van der Lugt, Aad
               and Verhulst, Frank C and Tiemeier, Henning and White, Tonya",
  journal   = "Neuropsychopharmacology",
  publisher = "Springer Science and Business Media LLC",
  volume    =  39,
  number    =  4,
  pages     = "792--800",
  month     =  mar,
  year      =  2014,
  language  = "en"
}

@article{gyrification_1,
author = {Madan, Christopher R.},
title = {Age-related decrements in cortical gyrification: Evidence from an accelerated longitudinal dataset},
journal = {European Journal of Neuroscience},
volume = {53},
number = {5},
pages = {1661-1671},
doi = {https://doi.org/10.1111/ejn.15039},
url = {https://onlinelibrary.wiley.com/doi/abs/10.1111/ejn.15039},
year = {2021}
}

@article{watanabe2017functional,
  title={Functional mapping and annotation of genetic associations with FUMA},
  author={Watanabe, Kyoko and Taskesen, Erdogan and Van Bochoven, Arjen and Posthuma, Danielle},
  journal={Nature communications},
  volume={8},
  number={1},
  pages={1826},
  year={2017},
  publisher={Nature Publishing Group UK London}
}

@ARTICLE{gyrification_2,
  title     = "Lifespan gyrification trajectories of human brain in healthy
               individuals and patients with major psychiatric disorders",
  author    = "Cao, Bo and Mwangi, Benson and Passos, Ives Cavalcante and Wu,
               Mon-Ju and Keser, Zafer and Zunta-Soares, Giovana B and Xu,
               Dianping and Hasan, Khader M and Soares, Jair C",
  journal   = "Sci. Rep.",
  publisher = "Springer Science and Business Media LLC",
  volume    =  7,
  number    =  1,
  pages     = "511",
  month     =  mar,
  year      =  2017,
  copyright = "https://creativecommons.org/licenses/by/4.0",
  language  = "en"
}

@article{van2020understanding,
  title={Understanding the genetic determinants of the brain with MOSTest},
  author={Van Der Meer, Dennis and Frei, Oleksandr and Kaufmann, Tobias and Shadrin, Alexey A and Devor, Anna and Smeland, Olav B and Thompson, Wesley K and Fan, Chun Chieh and Holland, Dominic and Westlye, Lars T and others},
  journal={Nature communications},
  volume={11},
  number={1},
  pages={3512},
  year={2020},
  publisher={Nature Publishing Group UK London}
}

@INPROCEEDINGS{JEPA,
  author={Assran, Mahmoud and Duval, Quentin and Misra, Ishan and Bojanowski, Piotr and Vincent, Pascal and Rabbat, Michael and LeCun, Yann and Ballas, Nicolas},
  booktitle={2023 IEEE/CVF Conference on Computer Vision and Pattern Recognition (CVPR)}, 
  title={Self-Supervised Learning from Images with a Joint-Embedding Predictive Architecture}, 
  year={2023},
  volume={},
  number={},
  pages={15619-15629},
  doi={10.1109/CVPR52729.2023.01499}}

@InProceedings{Caron_2021_ICCV,
    author    = {Caron, Mathilde and Touvron, Hugo and Misra, Ishan and J\'egou, Herv\'e and Mairal, Julien and Bojanowski, Piotr and Joulin, Armand},
    title     = {Emerging Properties in Self-Supervised Vision Transformers},
    booktitle = {Proceedings of the IEEE/CVF International Conference on Computer Vision (ICCV)},
    month     = {10},
    year      = {2021},
    pages     = {9650-9660}
}

@inproceedings{byol,
 author = {Grill, Jean-Bastien and Strub, Florian and Altch\'{e}, Florent and Tallec, Corentin and Richemond, Pierre and Buchatskaya, Elena and Doersch, Carl and Avila Pires, Bernardo and Guo, Zhaohan and Gheshlaghi Azar, Mohammad and Piot, Bilal and kavukcuoglu, koray and Munos, Remi and Valko, Michal},
 booktitle = {Advances in Neural Information Processing Systems},
 editor = {H. Larochelle and M. Ranzato and R. Hadsell and M.F. Balcan and H. Lin},
 pages = {21271--21284},
 publisher = {Curran Associates, Inc.},
 title = {Bootstrap Your Own Latent - A New Approach to Self-Supervised Learning},
 url = {https://proceedings.neurips.cc/paper_files/paper/2020/file/f3ada80d5c4ee70142b17b8192b2958e-Paper.pdf},
 volume = {33},
 year = {2020}
}

@inproceedings{swav,
author = {Caron, Mathilde and Misra, Ishan and Mairal, Julien and Goyal, Priya and Bojanowski, Piotr and Joulin, Armand},
title = {Unsupervised learning of visual features by contrasting cluster assignments},
year = {2020},
isbn = {9781713829546},
publisher = {Curran Associates Inc.},
address = {Red Hook, NY, USA},
booktitle = {Proceedings of the 34th International Conference on Neural Information Processing Systems},
articleno = {831},
numpages = {13},
location = {Vancouver, BC, Canada},
series = {NIPS '20}
}

@article{zhou2021ibot,
  title={iBOT: Image BERT Pre-Training with Online Tokenizer},
  author={Zhou, Jinghao and Wei, Chen and Wang, Huiyu and Shen, Wei and Xie, Cihang and Yuille, Alan and Kong, Tao},
  journal={International Conference on Learning Representations (ICLR)},
  year={2022}
}

@INPROCEEDINGS{He2022-yv,
  title           = "Masked Autoencoders Are Scalable Vision Learners",
  booktitle       = "2022 {IEEE/CVF} Conference on Computer Vision and Pattern
                     Recognition ({CVPR})",
  author          = "He, Kaiming and Chen, Xinlei and Xie, Saining and Li,
                     Yanghao and Dollar, Piotr and Girshick, Ross",
  publisher       = "IEEE",
  pages           = "15979--15988",
  month           =  jun,
  year            =  2022,
  conference      = "2022 IEEE/CVF Conference on Computer Vision and Pattern
                     Recognition (CVPR)",
  location        = "New Orleans, LA, USA"
}

@inproceedings{
higgins2017betavae,
title={beta-{VAE}: Learning Basic Visual Concepts with a Constrained Variational Framework},
author={Irina Higgins and Loic Matthey and Arka Pal and Christopher Burgess and Xavier Glorot and Matthew Botvinick and Shakir Mohamed and Alexander Lerchner},
booktitle={International Conference on Learning Representations},
year={2017},
url={https://openreview.net/forum?id=Sy2fzU9gl}
}

@inproceedings{
assel2025jointembedding,
title={Joint\nobreakdash-Embedding vs Reconstruction: Provable Benefits of Latent Space Prediction for Self\nobreakdash-Supervised Learning},
author={Hugues Van Assel and Mark Ibrahim and Tommaso Biancalani and Aviv Regev and Randall Balestriero},
booktitle={The Thirty-ninth Annual Conference on Neural Information Processing Systems},
year={2025},
url={https://openreview.net/forum?id=UOaLsgn5wb}
}

@ARTICLE{Wu2025-iz,
  title     = "Reassessing asymmetry reduction in psychosis: Cingulate folding
               and gyrification covariance in patients with auditory
               hallucinations",
  author    = "Wu, Shun-Chin Jim and de Vareilles, H{\'e}lo{\"\i}se and
               Mitchell, Samantha C and Al-Manea, Atheer and Garrison, Jane and
               Mamalakis, Michail and Simons, Jon S and Cachia, Arnaud and
               Mangin, Jean-Fran{\c c}ois and Nerland, Stener and
               M{\o}rch-Johnsen, Lynn and Agartz, Ingrid and Suckling, John and
               Murray, Graham K",
  journal   = "Schizophr. Bull.",
  publisher = "Oxford University Press (OUP)",
  month     =  jun,
  year      =  2025,
  copyright = "https://creativecommons.org/licenses/by/4.0/",
  language  = "en"
}

@ARTICLE{Cachia2008-go,
  title     = "Cortical folding abnormalities in schizophrenia patients with
               resistant auditory hallucinations",
  author    = "Cachia, Arnaud and Paill{\`e}re-Martinot, Marie-Laure and
               Galinowski, Andr{\'e} and Januel, Dominique and de Beaurepaire,
               Renaud and Bellivier, Frank and Artiges, Eric and Andoh, Jamila
               and Bartr{\'e}s-Faz, David and Duchesnay, Edouard and
               Rivi{\`e}re, Denis and ze, Marion and Mangin, Jean-Francois
               and Martinot, Jean-Luc",
  journal   = "Neuroimage",
  publisher = "Elsevier BV",
  volume    =  39,
  number    =  3,
  pages     = "927--935",
  month     =  feb,
  year      =  2008,
  language  = "en"
}

@ARTICLE{Penttila2008-pm,
  title     = "Global and temporal cortical folding in patients with
               early-onset schizophrenia",
  author    = "Penttil{\"a}, Jani and Paill{\'e}re-Martinot, Marie-Laure and
               Martinot, Jean-Luc and Mangin, Jean-Fran{\c c}ois and Burke,
               Lisa and Corrigall, Richard and Frangou, Sophia and Cachia,
               Arnaud",
  journal   = "J. Am. Acad. Child Adolesc. Psychiatry",
  publisher = "Elsevier BV",
  volume    =  47,
  number    =  10,
  pages     = "1125--1132",
  month     =  oct,
  year      =  2008,
  language  = "en"
}

@ARTICLE{Yucel2002-kt,
  title     = "Paracingulate morphologic differences in males with established
               schizophrenia: a magnetic resonance imaging morphometric study",
  author    = "Y{\"u}cel, Murat and Stuart, Geoffrey W and Maruff, Paul and
               Wood, Stephen J and Savage, Greg R and Smith, Deidre J and
               Crowe, Simon F and Copolov, David L and Velakoulis, Dennis and
               Pantelis, Christos",
  journal   = "Biol. Psychiatry",
  publisher = "Elsevier BV",
  volume    =  52,
  number    =  1,
  pages     = "15--23",
  month     =  jul,
  year      =  2002,
  language  = "en"
}

@ARTICLE{Provost2003-ux,
  title     = "Paracingulate sulcus morphology in men with early-onset
               schizophrenia",
  author    = "Provost, Jean-Bernard Le and Bartr{\'e}s-Faz, David and
               Paill{\`e}re-Martinot, Marie-Laure and Artiges, Eric and
               Pappata, Sabina and Recasens, Christophe and
               P{\'e}rez-G{\'o}mez, Mercedes and Bernardo, Miquel and Baeza,
               Imma and Bayle, Frank and Martinot, Jean-Luc",
  journal   = "Br. J. Psychiatry",
  publisher = "Royal College of Psychiatrists",
  volume    =  182,
  number    =  03,
  pages     = "228--232",
  month     =  mar,
  year      =  2003
}

@ARTICLE{Fujiwara2007-bo,
  title     = "Anterior cingulate pathology and social cognition in
               schizophrenia: a study of gray matter, white matter and sulcal
               morphometry",
  author    = "Fujiwara, Hironobu and Hirao, Kazuyuki and Namiki, Chihiro and
               Yamada, Makiko and Shimizu, Mitsuaki and Fukuyama, Hidenao and
               Hayashi, Takuji and Murai, Toshiya",
  journal   = "Neuroimage",
  publisher = "Elsevier BV",
  volume    =  36,
  number    =  4,
  pages     = "1236--1245",
  month     =  jul,
  year      =  2007,
  language  = "en"
}

@ARTICLE{Nakamura2007-pe,
  title     = "Altered orbitofrontal sulcogyral pattern in schizophrenia",
  author    = "Nakamura, Motoaki and Nestor, Paul G and McCarley, Robert W and
               Levitt, James J and Hsu, Lillian and Kawashima, Toshiro and
               Niznikiewicz, Margaret and Shenton, Martha E",
  journal   = "Brain",
  publisher = "Oxford University Press (OUP)",
  volume    =  130,
  number    = "Pt 3",
  pages     = "693--707",
  month     =  mar,
  year      =  2007,
  language  = "en"
}

@ARTICLE{Plaze2011-mk,
  title     = "``Where do auditory hallucinations come from?''--a brain
               morphometry study of schizophrenia patients with inner or outer
               space hallucinations",
  author    = "Plaze, Marion and Paill{\`e}re-Martinot, Marie-Laure and
               Penttil{\"a}, Jani and Januel, Dominique and de Beaurepaire,
               Renaud and Bellivier, Franck and Andoh, Jamila and Galinowski,
               Andr{\'e} and Gallarda, Thierry and Artiges, Eric and Oli{\'e},
               Jean-Pierre and Mangin, Jean-Fran{\c c}ois and Martinot,
               Jean-Luc and Cachia, Arnaud",
  journal   = "Schizophr. Bull.",
  publisher = "Oxford University Press (OUP)",
  volume    =  37,
  number    =  1,
  pages     = "212--221",
  month     =  jan,
  year      =  2011,
  language  = "en"
}

@ARTICLE{De_Juan_Romero2017-lr,
  title     = "Genetic maps and patterns of cerebral cortex folding",
  author    = "de Juan Romero, Camino and Borrell, V{\'\i}ctor",
  journal   = "Curr. Opin. Cell Biol.",
  publisher = "Elsevier BV",
  volume    =  49,
  pages     = "31--37",
  month     =  dec,
  year      =  2017,
  language  = "en"
}

@ARTICLE{Llinares-Benadero2019-jf,
  title     = "Deconstructing cortical folding: genetic, cellular and
               mechanical determinants",
  author    = "Llinares-Benadero, Cristina and Borrell, V{\'\i}ctor",
  journal   = "Nat. Rev. Neurosci.",
  publisher = "Springer Science and Business Media LLC",
  volume    =  20,
  number    =  3,
  pages     = "161--176",
  month     =  mar,
  year      =  2019,
  language  = "en"
}

@ARTICLE{Mangin2016-je,
  title     = "Spatial normalization of brain images and beyond",
  author    = "Mangin, J-F and Lebenberg, J and Lefranc, S and Labra, N and
               Auzias, G and Labit, M and Guevara, M and Mohlberg, H and Roca,
               P and Guevara, P and Dubois, J and Leroy, F and
               Dehaene-Lambertz, G and Cachia, A and Dickscheid, T and Coulon,
               O and Poupon, C and Rivi{\`e}re, D and Amunts, K and Sun, Z Y",
  journal   = "Med. Image Anal.",
  publisher = "Elsevier BV",
  volume    =  33,
  pages     = "127--133",
  month     =  oct,
  year      =  2016
}

@ARTICLE{De_Vareilles2023-ri,
  title     = "Development of cortical folds in the human brain: An attempt to
               review biological hypotheses, early neuroimaging investigations
               and functional correlates",
  author    = "de Vareilles, H and Rivi{\`e}re, D and Mangin, J F and Dubois, J",
  journal   = "Dev. Cogn. Neurosci.",
  publisher = "Elsevier BV",
  volume    =  61,
  number    =  101249,
  pages     = "101249",
  month     =  jun,
  year      =  2023,
  copyright = "http://creativecommons.org/licenses/by-nc-nd/4.0/",
  language  = "en"
}

@ARTICLE{Snyder2024,
  title     = "A bimodal taxonomy of adult human brain sulcal morphology
               related to timing of fetal sulcation and trans-sulcal gene
               expression gradients",
  author    = "Snyder, William E and V{\'e}rtes, Petra E and Kyriakopoulou,
               Vanessa and Wagstyl, Konrad and Williams, Logan Z J and
               Moraczewski, Dustin and Thomas, Adam G and Karolis, Vyacheslav R
               and Seidlitz, Jakob and Rivi{\`e}re, Denis and Robinson, Emma C
               and Mangin, Jean-Francois and Raznahan, Armin and Bullmore,
               Edward T",
  journal   = "Neuron",
  publisher = "Elsevier BV",
  volume    =  112,
  number    =  20,
  pages     = "3396--3411.e6",
  month     =  oct,
  year      =  2024,
  copyright = "http://creativecommons.org/licenses/by/4.0/",
  language  = "en"
}

@ARTICLE{Guillon2024,
  title     = "Identification of rare cortical folding patterns using
               unsupervised deep learning",
  author    = "Guillon, Louise and Chavas, Jo{\"e}l and B{\'e}n{\'e}zit, Audrey
               and Moutard, Marie-Laure and Roca, Pauline and Mellerio, Charles
               and Oppenheim, Catherine and Rivi{\`e}re, Denis and Mangin,
               Jean-Fran{\c c}ois",
  journal   = "Imaging Neurosci. (Camb.)",
  publisher = "MIT Press",
  volume    =  2,
  pages     = "1--27",
  month     =  feb,
  year      =  2024,
  copyright = "https://creativecommons.org/licenses/by/4.0/",
  language  = "en"
}

@InProceedings{cka,
  title = 	 {Similarity of Neural Network Representations Revisited},
  author =       {Kornblith, Simon and Norouzi, Mohammad and Lee, Honglak and Hinton, Geoffrey},
  booktitle = 	 {Proceedings of the 36th International Conference on Machine Learning},
  pages = 	 {3519--3529},
  year = 	 {2019},
  editor = 	 {Chaudhuri, Kamalika and Salakhutdinov, Ruslan},
  volume = 	 {97},
  series = 	 {Proceedings of Machine Learning Research},
  month = 	 {09--15 Jun},
  publisher =    {PMLR},
  url = 	 {https://proceedings.mlr.press/v97/kornblith19a.html}
}

@article{scikit-learn,
  title={Scikit-learn: Machine Learning in {P}ython},
  author={Pedregosa, F. and Varoquaux, G. and Gramfort, A. and Michel, V.
          and Thirion, B. and Grisel, O. and Blondel, M. and Prettenhofer, P.
          and Weiss, R. and Dubourg, V. and Vanderplas, J. and Passos, A. and
          Cournapeau, D. and Brucher, M. and Perrot, M. and Duchesnay, E.},
  journal={Journal of Machine Learning Research},
  volume={12},
  pages={2825--2830},
  year={2011}
}

@article{Santacroce2024-dv,
  title     = "Human intraparietal sulcal morphology relates to individual
               differences in language and memory performance",
  author    = "Santacroce, Federica and Cachia, Arnaud and Fragueiro, Agustina
               and Grande, Eleonora and Roell, Margot and Baldassarre,
               Antonello and Sestieri, Carlo and Committeri, Giorgia",
  journal   = "Commun. Biol.",
  publisher = "Springer Science and Business Media LLC",
  volume    =  7,
  number    =  1,
  pages     = "520",
  month     =  may,
  year      =  2024,
  copyright = "https://creativecommons.org/licenses/by/4.0",
  language  = "en"
}

@article{cachia_shape_2014,
	title = {The shape of the {ACC} contributes to cognitive control efficiency in preschoolers},
	volume = {26},
	issn = {1530-8898},
	doi = {10.1162/jocn_a_00459},
	language = {eng},
	number = {1},
	journal = {Journal of Cognitive Neuroscience},
	author = {Cachia, Arnaud and Borst, Grégoire and Vidal, Julie and Fischer, Clara and Pineau, Arlette and Mangin, Jean-François and Houdé, Olivier},
	month = jan,
	year = {2014},
	pmid = {23915057},
	pages = {96--106},
}

@article{moyal_orbitofrontal_2024,
	title = {Orbitofrontal sulcal patterns in catatonia},
	volume = {67},
	issn = {0924-9338, 1778-3585},
	url = {https://www.cambridge.org/core/journals/european-psychiatry/article/orbitofrontal-sulcal-patterns-in-catatonia/BB55D396F8243A3EE6DE5F2ED1DB6108},
	doi = {10.1192/j.eurpsy.2023.2461},
	language = {en},
	number = {1},
	journal = {European Psychiatry},
	author = {Moyal, Mylène and Haroche, Alexandre and Attali, David and Dadi, Ghita and Raoelison, Matthieu and Berre, Alice Le and Iftimovici, Anton and Chaumette, Boris and Leroy, Sylvain and Charron, Sylvain and Debacker, Clément and Oppenheim, Catherine and Cachia, Arnaud and Plaze, Marion},
	month = jan,
	year = {2024},
	pages = {e6},
}

@article{isomura_altered_2017,
	title = {Altered sulcogyral patterns of orbitofrontal cortex in a large cohort of patients with schizophrenia},
	volume = {3},
	issn = {2334-265X},
	doi = {10.1038/s41537-016-0008-y},
	language = {eng},
	journal = {NPJ schizophrenia},
	author = {Isomura, Shuichi and Hashimoto, Ryota and Nakamura, Motoaki and Hirano, Yoji and Yamashita, Fumio and Jimbo, Shin and Yamamori, Hidenaga and Fujimoto, Michiko and Yasuda, Yuka and Mears, Ryan P. and Onitsuka, Toshiaki},
	year = {2017},
	pmid = {28560249},
	pmcid = {PMC5441528},
	pages = {3},
}

@article{bycroft_uk_2018,
	title = {The {UK} {Biobank} resource with deep phenotyping and genomic data},
	volume = {562},
	copyright = {2018 Springer Nature Limited},
	issn = {1476-4687},
	url = {https://www.nature.com/articles/s41586-018-0579-z},
	doi = {10.1038/s41586-018-0579-z},
	language = {en},
	number = {7726},
	journal = {Nature},
	author = {Bycroft, Clare and Freeman, Colin and Petkova, Desislava and Band, Gavin and Elliott, Lloyd T. and Sharp, Kevin and Motyer, Allan and Vukcevic, Damjan and Delaneau, Olivier and O’Connell, Jared and Cortes, Adrian and Welsh, Samantha and Young, Alan and Effingham, Mark and McVean, Gil and Leslie, Stephen and Allen, Naomi and Donnelly, Peter and Marchini, Jonathan},
	month = oct,
	year = {2018},
	note = {Publisher: Nature Publishing Group},
	pages = {203--209},
}

@article{van_essen_wu-minn_2013,
	series = {Mapping the {Connectome}},
	title = {The {WU}-{Minn} {Human} {Connectome} {Project}: {An} overview},
	volume = {80},
	issn = {1053-8119},
	shorttitle = {The {WU}-{Minn} {Human} {Connectome} {Project}},
	url = {https://www.sciencedirect.com/science/article/pii/S1053811913005351},
	doi = {10.1016/j.neuroimage.2013.05.041},
	journal = {NeuroImage},
	author = {Van Essen, David C. and Smith, Stephen M. and Barch, Deanna M. and Behrens, Timothy E. J. and Yacoub, Essa and Ugurbil, Kamil},
	month = oct,
	year = {2013},
	pages = {62--79}
}

@article{tenenbaum_global_2000,
	title = {A {Global} {Geometric} {Framework} for {Nonlinear} {Dimensionality} {Reduction}},
	volume = {290},
	issn = {0036-8075, 1095-9203},
	url = {https://www.science.org/doi/10.1126/science.290.5500.2319},
	doi = {10.1126/science.290.5500.2319},
	language = {en},
	number = {5500},
	journal = {Science},
	author = {Tenenbaum, Joshua B. and Silva, Vin De and Langford, John C.},
	month = dec,
	year = {2000},
	pages = {2319--2323},
}

@article{chakravarty_striatal_2014,
	title = {Striatal shape abnormalities as novel neurodevelopmental endophenotypes in schizophrenia: {A} longitudinal study},
	volume = {36},
	issn = {1065-9471},
	shorttitle = {Striatal shape abnormalities as novel neurodevelopmental endophenotypes in schizophrenia},
	url = {https://www.ncbi.nlm.nih.gov/pmc/articles/PMC6869651/},
	doi = {10.1002/hbm.22715},
	number = {4},
	journal = {Human Brain Mapping},
	author = {Chakravarty, M. Mallar and Rapoport, Judith L. and Giedd, Jay N. and Raznahan, Armin and Shaw, Philip and Collins, D. Louis and Lerch, Jason P. and Gogtay, Nitin},
	month = dec,
	year = {2014},
	pmid = {25504933},
	pmcid = {PMC6869651},
	pages = {1458--1469},
}

@article{rapoport_childhood_2011,
	title = {Childhood onset schizophrenia: support for a progressive neurodevelopmental disorder},
	volume = {29},
	issn = {1873-474X},
	shorttitle = {Childhood onset schizophrenia},
	doi = {10.1016/j.ijdevneu.2010.10.003},
	language = {eng},
	number = {3},
	journal = {International Journal of Developmental Neuroscience: The Official Journal of the International Society for Developmental Neuroscience},
	author = {Rapoport, Judith L. and Gogtay, Nitin},
	month = may,
	year = {2011},
	pmid = {20955775},
	pmcid = {PMC5157162},
	pages = {251--258},
}

@article{cachia_longitudinal_2016,
	title = {Longitudinal stability of the folding pattern of the anterior cingulate cortex during development},
	volume = {19},
	issn = {1878-9307},
	doi = {10.1016/j.dcn.2016.02.011},
	language = {eng},
	journal = {Developmental Cognitive Neuroscience},
	author = {Cachia, A. and Borst, G. and Tissier, C. and Fisher, C. and Plaze, M. and Gay, O. and Rivière, D. and Gogtay, N. and Giedd, J. and Mangin, J.-F. and Houdé, O. and Raznahan, A.},
	month = jun,
	year = {2016},
	pmid = {26974743},
	pmcid = {PMC4912935},
	pages = {122--127},
}

@article{delalande_complex_2020,
	title = {Complex and subtle structural changes in prefrontal cortex induced by inhibitory control training from childhood to adolescence},
	volume = {23},
	issn = {1467-7687},
	doi = {10.1111/desc.12898},
	language = {eng},
	number = {4},
	journal = {Developmental Science},
	author = {Delalande, Lisa and Moyon, Marine and Tissier, Cloélia and Dorriere, Valérie and Guillois, Bernard and Mevell, Katel and Charron, Sylvain and Salvia, Emilie and Poirel, Nicolas and Vidal, Julie and Lion, Stéphanie and Oppenheim, Catherine and Houdé, Olivier and Cachia, Arnaud and Borst, Grégoire},
	month = jul,
	year = {2020},
	pmid = {31469938},
	pages = {e12898},
}

@article{tissier_sulcal_2018,
	title = {Sulcal {Polymorphisms} of the {IFC} and {ACC} {Contribute} to {Inhibitory} {Control} {Variability} in {Children} and {Adults}},
	volume = {5},
	issn = {2373-2822},
	doi = {10.1523/ENEURO.0197-17.2018},
	language = {eng},
	number = {1},
	journal = {eNeuro},
	author = {Tissier, Cloélia and Linzarini, Adriano and Allaire-Duquette, Geneviève and Mevel, Katell and Poirel, Nicolas and Dollfus, Sonia and Etard, Olivier and Orliac, François and Peyrin, Carole and Charron, Sylvain and Raznahan, Armin and Houdé, Olivier and Borst, Grégoire and Cachia, Arnaud},
	year = {2018},
	pmid = {29527565},
	pmcid = {PMC5844057},
	pages = {ENEURO.0197--17.2018},
}

@article{sun_linking_2016,
	title = {Linking morphological and functional variability in hand movement and silent reading},
	volume = {221},
	issn = {1863-2661},
	doi = {10.1007/s00429-015-1106-8},
	language = {eng},
	number = {7},
	journal = {Brain Structure \& Function},
	author = {Sun, Z. Y. and Pinel, P. and Rivière, D. and Moreno, A. and Dehaene, S. and Mangin, J.-F.},
	month = sep,
	year = {2016},
	pmid = {26346119},
	pages = {3361--3371},
}

@article{chen_simple_2020,
  author       = {Ting Chen and
                  Simon Kornblith and
                  Mohammad Norouzi and
                  Geoffrey E. Hinton},
  title        = {A Simple Framework for Contrastive Learning of Visual Representations},
  journal      = {International conference on machine learning},
  publisher =    {PMLR},
  pages        = {1597–1607},
  volume       = {119},
  year         = {2020},
  url          = 	 {https://proceedings.mlr.press/v119/chen20j.html},
}

@InProceedings{zbontar_barlow_2021,
  title = 	 {Barlow Twins: Self-Supervised Learning via Redundancy Reduction},
  author =       {Zbontar, Jure and Jing, Li and Misra, Ishan and LeCun, Yann and Deny, Stephane},
  booktitle = 	 {Proceedings of the 38th International Conference on Machine Learning},
  pages = 	 {12310--12320},
  year = 	 {2021},
  editor = 	 {Meila, Marina and Zhang, Tong},
  volume = 	 {139},
  series = 	 {Proceedings of Machine Learning Research},
  month = 	 {18--24 Jul},
  publisher =    {PMLR},
  url = 	 {https://proceedings.mlr.press/v139/zbontar21a.html}
}

@article{balestriero_cookbook_2023,
  title={A cookbook of self-supervised learning},
  author={Balestriero, Randall and Ibrahim, Mark and Sobal, Vlad and Morcos, Ari and Shekhar, Shashank and Goldstein, Tom and Bordes, Florian and Bardes, Adrien and Mialon, Gregoire and Tian, Yuandong and others},
  journal={arXiv preprint arXiv:2304.12210},
  year={2023}
}

@article{troiani_variability_2022,
	title = {Variability and concordance of sulcal patterns in the orbitofrontal cortex: {A} twin study},
	volume = {324},
	issn = {0925-4927},
	shorttitle = {Variability and concordance of sulcal patterns in the orbitofrontal cortex},
	url = {https://www.sciencedirect.com/science/article/pii/S0925492722000531},
	doi = {10.1016/j.pscychresns.2022.111492},
	journal = {Psychiatry Research: Neuroimaging},
	author = {Troiani, Vanessa and Snyder, Will and Kozick, Shane and Patti, Marisa A and Beiler, Donielle},
	month = aug,
	year = {2022},
	pages = {111492},
}

@article{dufumier_exploring_2024,
title = {Exploring the potential of representation and transfer learning for anatomical neuroimaging: Application to psychiatry},
journal = {NeuroImage},
volume = {296},
pages = {120665},
year = {2024},
issn = {1053-8119},
doi = {https://doi.org/10.1016/j.neuroimage.2024.120665},
url = {https://www.sciencedirect.com/science/article/pii/S1053811924001605},
author = {Benoit Dufumier and Pietro Gori and Sara Petiton and Robin Louiset and Jean-François Mangin and Antoine Grigis and Edouard Duchesnay},
}

@misc{devries_improved_2017,
	title = {Improved {Regularization} of {Convolutional} {Neural} {Networks} with {Cutout}},
	url = {http://arxiv.org/abs/1708.04552},
	language = {en},
	publisher = {arXiv},
	author = {DeVries, Terrance and Taylor, Graham W.},
	month = nov,
	year = {2017},
	note = {arXiv:1708.04552 [cs]},
}

@INPROCEEDINGS{resnet,
  author={He, Kaiming and Zhang, Xiangyu and Ren, Shaoqing and Sun, Jian},
  booktitle={2016 IEEE Conference on Computer Vision and Pattern Recognition (CVPR)}, 
  title={Deep Residual Learning for Image Recognition}, 
  year={2016},
  volume={},
  number={},
  pages={770-778},
  doi={10.1109/CVPR.2016.90}}

@article{sun_effect_2012,
	title = {The effect of handedness on the shape of the central sulcus},
	volume = {60},
	issn = {1053-8119},
	url = {https://www.sciencedirect.com/science/article/pii/S1053811911014522},
	doi = {10.1016/j.neuroimage.2011.12.050},
	number = {1},
	journal = {NeuroImage},
	author = {Sun, Zhong Yi and Klöppel, Stefan and Rivière, Denis and Perrot, Matthieu and Frackowiak, Richard and Siebner, Hartwig and Mangin, Jean-François},
	month = mar,
	year = {2012},
	pages = {332--339},
}

@inproceedings{ter_braak_permutation_1992,
	address = {Berlin, Heidelberg},
	title = {Permutation {Versus} {Bootstrap} {Significance} {Tests} in {Multiple} {Regression} and {Anova}},
	isbn = {978-3-642-48850-4},
	doi = {10.1007/978-3-642-48850-4_10},
	language = {en},
	booktitle = {Bootstrapping and {Related} {Techniques}},
	publisher = {Springer},
	author = {ter Braak, Cajo J. F.},
	editor = {Jöckel, Karl-Heinz and Rothe, Günter and Sendler, Wolfgang},
	year = {1992},
	pages = {79--85},
}

@article{de_matos_temporo-basal_2023,
	title = {Temporo-basal sulcal connections: a manual annotation protocol and an investigation of sexual dimorphism and heritability},
	volume = {228},
	issn = {1863-2661},
	shorttitle = {Temporo-basal sulcal connections},
	url = {https://link.springer.com/10.1007/s00429-023-02663-6},
	doi = {10.1007/s00429-023-02663-6},
	language = {en},
	number = {6},
	journal = {Brain Structure and Function},
	author = {De Matos, Kevin and Cury, Claire and Chougar, Lydia and Strike, Lachlan T. and Rolland, Thibault and Riche, Maximilien and Hemforth, Lisa and Martin, Alexandre and Banaschewski, Tobias and Bokde, Arun L. W. and Desrivières, Sylvane and Flor, Herta and Grigis, Antoine and Garavan, Hugh and Gowland, Penny and Heinz, Andreas and Brühl, Rüdiger and Martinot, Jean-Luc and Paillère Martinot, Marie-Laure and Artiges, Eric and Nees, Frauke and Papadopoulos Orfanos, Dimitri and Lemaitre, Herve and Paus, Tomáš and Poustka, Luise and Hohmann, Sarah and Millenet, Sabina and Fröhner, Juliane H. and Smolka, Michael N. and Vaidya, Nilakshi and Walter, Henrik and Whelan, Robert and Schumann, Gunter and Frouin, Vincent and {IMAGEN Consortium} and Bach Cuadra, Meritxell and Colliot, Olivier and Couvy-Duchesne, Baptiste},
	month = jun,
	year = {2023},
	pages = {1459--1478},
}

@article{fortin2018harmonization,
  title={Harmonization of cortical thickness measurements across scanners and sites},
  author={Fortin, Jean-Philippe and Cullen, Nicholas and Sheline, Yvette I and Taylor, Warren D and Aselcioglu, Irem and Cook, Philip A and Adams, Phil and Cooper, Crystal and Fava, Maurizio and McGrath, Patrick J and others},
  journal={Neuroimage},
  volume={167},
  pages={104--120},
  year={2018},
  publisher={Elsevier}
}

@article{johnson2007adjusting,
  title={Adjusting batch effects in microarray expression data using empirical Bayes methods},
  author={Johnson, W Evan and Li, Cheng and Rabinovic, Ariel},
  journal={Biostatistics},
  volume={8},
  number={1},
  pages={118--127},
  year={2007},
  publisher={Oxford University Press}
}

@dataset{qtim,
  author = {Strike, Lachlan T. AND Blokland, Gabriella A.M. AND Hansell, Narelle K. AND Martin, Nicholas G. AND Toga, Arthur W. AND Thompson, Paul M. AND de Zubicaray, Greig I. AND McMahon, Katie L. AND Wright, Margaret J.},
  title = {"Queensland Twin IMaging (QTIM)"},
  year = {2022},
  doi = {doi:10.18112/openneuro.ds004169.v1.0.6},
  publisher = {OpenNeuro}
}

@article{strike_genetic_2019,
	title = {Genetic {Complexity} of {Cortical} {Structure}: {Differences} in {Genetic} and {Environmental} {Factors} {Influencing} {Cortical} {Surface} {Area} and {Thickness}},
	volume = {29},
	issn = {1047-3211},
	shorttitle = {Genetic {Complexity} of {Cortical} {Structure}},
	url = {https://doi.org/10.1093/cercor/bhy002},
	doi = {10.1093/cercor/bhy002},
	number = {3},
	journal = {Cerebral Cortex},
	author = {Strike, Lachlan T and Hansell, Narelle K and Couvy-Duchesne, Baptiste and Thompson, Paul M and de Zubicaray, Greig I and McMahon, Katie L and Wright, Margaret J},
	month = mar,
	year = {2019},
	pages = {952--962},
}

@article{volkow2018conception,
  title={The conception of the ABCD study: From substance use to a broad NIH collaboration},
  author={Volkow, Nora D and Koob, George F and Croyle, Robert T and Bianchi, Diana W and Gordon, Joshua A and Koroshetz, Walter J and P{\'e}rez-Stable, Eliseo J and Riley, William T and Bloch, Michele H and Conway, Kevin and others},
  journal={Developmental cognitive neuroscience},
  volume={32},
  pages={4--7},
  year={2018},
  publisher={Elsevier}
}

@article{sudlow2015uk,
  title={UK biobank: an open access resource for identifying the causes of a wide range of complex diseases of middle and old age},
  author={Sudlow, Cathie and Gallacher, John and Allen, Naomi and Beral, Valerie and Burton, Paul and Danesh, John and Downey, Paul and Elliott, Paul and Green, Jane and Landray, Martin and others},
  journal={PLoS medicine},
  volume={12},
  number={3},
  pages={e1001779},
  year={2015},
  publisher={Public Library of Science}
}

@article{cury2015incomplete,
  title={Incomplete hippocampal inversion: a comprehensive MRI study of over 2000 subjects},
  author={Cury, Claire and Toro, Roberto and Cohen, Fanny and Fischer, Clara and Mhaya, Amel and Samper-Gonz{\'a}lez, Jorge and Hasboun, Dominique and Mangin, Jean-Fran{\c{c}}ois and Banaschewski, Tobias and Bokde, Arun LW and others},
  journal={Frontiers in neuroanatomy},
  volume={9},
  pages={160},
  year={2015},
  publisher={Frontiers Media SA}
}

@article{derauf_subcortical_2012,
	title = {Subcortical and {Cortical} {Structural} {Central} {Nervous} {System} {Changes} and {Attention} {Processing} {Deficits} in {Preschool}-{Aged} {Children} with {Prenatal} {Methamphetamine} and {Tobacco} {Exposure}},
	volume = {34},
	issn = {0378-5866},
	url = {https://pmc.ncbi.nlm.nih.gov/articles/PMC4091037/},
	doi = {10.1159/000341119},
	number = {4},
	journal = {Developmental neuroscience},
	author = {Derauf, Chris and Lester, Barry M. and Neyzi, Nurunisa and Kekatpure, Minal and Gracia, Luis and Davis, James and Kallianpur, Kalpana and Efird, Jimmy T. and Kosofsky, Barry},
	year = {2012},
	pages = {327--341},
}

@article{voorhies_cognitive_2021,
	title = {Cognitive insights from tertiary sulci in prefrontal cortex},
	volume = {12},
	copyright = {2021 The Author(s)},
	issn = {2041-1723},
	url = {https://www.nature.com/articles/s41467-021-25162-w},
	doi = {10.1038/s41467-021-25162-w},
	language = {en},
	number = {1},
	journal = {Nature Communications},
	publisher = {Nature Publishing Group},
	author = {Voorhies, Willa I. and Miller, Jacob A. and Yao, Jewelia K. and Bunge, Silvia A. and Weiner, Kevin S.},
	month = aug,
	year = {2021},
	pages = {5122},
}

@article{liang_cortical_2025,
	title = {A cortical signature of very preterm birth across development and its association with neurodevelopmental outcomes},
	url = {https://www.biorxiv.org/content/10.1101/2025.10.17.683086.abstract},
	journal = {bioRxiv},
	publisher = {Cold Spring Harbor Laboratory},
	author = {Liang, Kaili and Guo, Yourong and Williams, Logan ZJ and Besenczi, Renato and Sun, Zeyuan and David Edwards, A. and Robinson, Emma C. and Nosarti, Chiara},
	year = {2025},
	pages = {2025--10},
}

@article{garrison_paracingulate_2015,
	title = {Paracingulate sulcus morphology is associated with hallucinations in the human brain},
	volume = {6},
	copyright = {2015 The Author(s)},
	issn = {2041-1723},
	url = {https://www.nature.com/articles/ncomms9956},
	doi = {10.1038/ncomms9956},
	language = {en},
	number = {1},
	journal = {Nature Communications},
	publisher = {Nature Publishing Group},
	author = {Garrison, Jane R. and Fernyhough, Charles and McCarthy-Jones, Simon and Haggard, Mark and Simons, Jon S.},
	month = nov,
	year = {2015},
	pages = {8956},
}

@article{rollins_evidence_2020,
	title = {Evidence in cortical folding patterns for prenatal predispositions to hallucinations in schizophrenia},
	volume = {10},
	copyright = {2020 The Author(s)},
	issn = {2158-3188},
	url = {https://www.nature.com/articles/s41398-020-01075-y},
	doi = {10.1038/s41398-020-01075-y},
	language = {en},
	number = {1},
	journal = {Translational Psychiatry},
	publisher = {Nature Publishing Group},
	author = {Rollins, Colleen P. E. and Garrison, Jane R. and Arribas, Maite and Seyedsalehi, Aida and Li, Zhi and Chan, Raymond C. K. and Yang, Junwei and Wang, Duo and Liò, Pietro and Yan, Chao and Yi, Zheng-hui and Cachia, Arnaud and Upthegrove, Rachel and Deakin, Bill and Simons, Jon S. and Murray, Graham K. and Suckling, John},
	month = nov,
	year = {2020},
	pages = {387},
}

\section{Acknowledgments}

This research was conducted using the \textbf{UK Biobank} resource under application number \textbf{64984}. This project has been funded by ANR via the Audace research program of the CEA, the IHU ICE (ANR-23-IAHU-0010), by AXA foundation for the project premaIA, by the "Fondation de France" for the project TSA-ideogrammes, by the PEPR Digital Health via BHT (ANR-22-PESN-0012),  by European Union’s Horizon 2020 for EBRAINS2 (HORIZON-INFRA-2022-SERV-B-01). This project was provided with computing HPC and storage resources by \textbf{GENCI TGCC}, thanks to the grant \textbf{2024-A0170313800} on the \textbf{Joliot Curie’s ROME} supercomputer partition. This project was provided with computer and storage resources by \textbf{GENCI at IDRIS} thanks to the grant \textbf{AD010316018} on the supercomputer Jean Zay’s V100 partition.

The research leading to these results has received funding from the French government under management of Agence Nationale de la Recherche as part of the “France 2030” program (reference ANR-23-IACL-0008, project PRAIRIE-PSAI), as part of the "Investissements d'avenir" program (reference ANR-10-IAIHU-06, project Agence Nationale de la Recherche-10-IA Institut Hospitalo-Universitaire-6), and from the European Union’s Horizon Europe Framework Programme (grant number 101136607, project CLARA).

We would like to thank Clément Langlet and Benoît Dufumier for their feedback on the paper, and Héloïse de Vareilles, Timothée Sanchez, and Théo Delmaire for their visualization ideas.
ChatGPT and Claude AI helped improve readability. \\

\section{Author contributions}

J. Laval, P. Gori, D. Rivière, J. Chavas, and J.-F. Mangin conceived the research idea and experiments. J. Laval wrote the manuscript, designed and optimized the model, conducted representation space analyses and interpretation, R. Guiavarch evaluated all foundation models. A. Dufournet conducted the genetic experiments and designed the decoder. R. Menasria developed the brain wide association pipeline. B. Drabczuk developed the final unified code. C. Mendoza generated visualizations. S. Tounsi and A. Baroud provided advice on exploratory analysis design, M. Bourenane developed scripts for cluster use, V Troiani, W. Snyder, M. Patti, M. Moyal, M. Plaze, A. Cachia, C. Cury, K. De Matos, O. Colliot, and Z.Y. Sun provided data and annotations, C. Fischer and V. Frouin preprocessed most of the data. C. Fisher and D. Rivière provided sulcal morphometry. J. Chavas developed the preprocessing and made preliminary work on the model design.
All authors read and approved the manuscript.

\section{Competing interests}

We declare no conflict of interest.

\clearpage

\section{Extended Data}

\begin{table}[!ht]
\centering
\small
\begin{tabular}{lC{1.3cm}C{1.5cm}C{1.5cm}C{1.cm}C{1.4cm}C{1.4cm}}
\toprule
Dataset & \#Subjects & Sex (\%F) & Age mean [min, max] & \#Sites & Magnetic Field (Tesla) & Resolution (isotropic)\\
\midrule
UK Biobank~\cite{bycroft_uk_2018} & 42,433 & 53 & 64 [44, 82] & 3 & 3T & 1mm\\
HCP~\cite{van_essen_wu-minn_2013} & 1,114 & 54 & 29 [22, 40] & 1 & 3T & 0.7mm\\
ACCpatterns~\cite{chakravarty_striatal_2014,rapoport_childhood_2011,cachia_longitudinal_2016,delalande_complex_2020,tissier_sulcal_2018} & 341 & 42 & 15 [8, 40] & 4 & 1.5T-3T & 1-2mm\\
ABCD~\cite{volkow2018conception} & 9,994 & 53 & 10 [9, 11] & 22 & 3T & 1mm\\
QTIM~\cite{qtim,strike_genetic_2019} & 928 & 61 & 21 [18, 30] & 1 & 4T & 0.9mm\\
\bottomrule
\end{tabular}
\caption{Overview of datasets. The UK Biobank was used for self-supervised learning (SSL) pretraining. All other datasets were employed exclusively in downstream analyses. This holds for Champollion as well as for the decoder. Although UK Biobank contains three sites, they are homogeneous (same scanner, same protocol). The ACCpatterns dataset is an aggregation of different datasets with different acquisition fields and resolutions. Some images are non-isotropic: images from~\cite{rapoport_childhood_2011} use 1.5-mm slices in the axial plane and 2.0-mm slices in the coronal plane.}
\label{tab:datasets}
\end{table}

\clearpage

\clearpage

\clearpage

\clearpage

\clearpage

\begin{table}[p]
\centering
\begin{tabular}{lr}
\hline
\textbf{BrainVISA Acronym} & \textbf{Sulcus / Fissure / Area name} \\
\hline
BROCA        & Broca's area \\
FCLa          & Anterior Sylvian fissure \\
FCLp          & Posterior Sylvian fissure \\
FCMant        & Anterior cingulate sulcus \\
FCMpost       & Posterior cingulate sulcus \\
FColl           & Collateral sulcus \\
FIP            & Intraparietal sulcus \\
FIPPoCinf   & Inferior postcentral sulcus \\
FPO            & Occipitoparietal sulcus \\
INSULA            & Insula \\
Lobule parietal sup & Postcentral, internal pariatal and intraparietal sulci  \\
OCCIPITAL         & Occipital sulci \\
SC              & Central sulcus \\
SCSylvian      & Central Sylvian sulcus \\
SCall           & Subcallosal sulcus \\
ScCal         & Calcarine sulcus \\
SCu             & Cuneal sulcus \\
SFinf          & Inferior frontal sulcus \\
SFinfant      & Anterior inferior frontal sulcus \\
SFint          & Internal frontal sulcus \\
SFinter        & Intermediate frontal sulcus \\
SFmarginal     & Marginal frontal sulcus \\
SFmedian       & Median frontal sulcus \\
SFpoltr         & Transverse polar frontal sulcus \\
SFsup          & Superior frontal sulcus \\
SintraCing      & Intracingulate sulcus \\
SLi             & Intralingual sulcus \\
SOTlat        & Lateral occipitotemporal sulcus \\
SOlf             & Olfactive sulcus \\
SOr             & Orbital sulcus \\
SpC            & Paracentral sulcus \\
SPaint         & Internal parietal sulcus \\
SPeC           & Precentral sulcus \\
SPeCinf       & Inferior precentral sulcus \\
SPoC           & Postcentral sulcus \\
SR              & Rostral sulcus \\
SRh             & Rhinal sulcus \\
SsP            & Sub-parietal sulcus \\
STi            & Inferior temporal sulcus \\
STpol          & Polar temporal sulcus \\
STs            & Superior temporal sulcus \\
STsbr            & Terminal ascending branches of the superior temporal sulcus \\
subsc          & Sub-central sulci \\
\hline
\end{tabular}
\caption{Acronyms for sulcus, fissure, and areas names. Region names used in this study are composed of the BrainVISA Acronyms.}
\label{tab:acronyms}
\end{table}

\clearpage

\clearpage

\end{document}